\documentclass[twocolumn,prb,aps,floatfix,longbibliography,superscriptaddress]{revtex4-2}

\usepackage{color}
\usepackage{graphicx}
\usepackage{siunitx}
\usepackage[color=green!60,textsize=small]{todonotes}
\usepackage{physics}
\usepackage{amsthm}
\usepackage{amsmath}
\usepackage{amssymb}
\usepackage{enumerate}
\usepackage{placeins}
\usepackage{booktabs}
\usepackage{dsfont}

\newif\ifarXiv
\arXivtrue 

\ifarXiv

\usepackage{pdfpages} 
\usepackage{pgffor} 
\usepackage{xr} 

\makeatletter
\AtBeginDocument{\let\LS@rot\@undefined}
\makeatother

\def\supplementfilename{supplementalScattering}

\pdfximage{\supplementfilename.pdf}
\def\numbersupplementpages{\the\pdflastximagepages}

\fi 

\usepackage{hyperref}
\hypersetup{
    colorlinks=true,
    linkcolor=blue,
    filecolor=magenta,      
    urlcolor=blue,
    citecolor=blue,
}

\renewcommand{\vec}[1]{\boldsymbol{#1}}

\AtBeginDocument{%
\heavyrulewidth=.08em
\lightrulewidth=.05em
\cmidrulewidth=.03em
\belowrulesep=.65ex
\belowbottomsep=0pt
\aboverulesep=.4ex
\abovetopsep=0pt
\cmidrulesep=\doublerulesep
\cmidrulekern=.5em
\defaultaddspace=.5em
}

\begin{document}

\title{Experimental Realization of One Dimensional Helium}

\author{Adrian Del Maestro}
\affiliation{Department of Physics and Astronomy, University of Tennessee, Knoxville, TN 37996, USA}
\affiliation{Min H. Kao Department of Electrical Engineering and Computer Science, University of Tennessee, Knoxville, TN 37996, USA}
\affiliation{Institute for Advanced Materials and Manufacturing, University of Tennessee, Knoxville, Tennessee 37996, USA\looseness=-1}

\author{Nathan S. Nichols}
\affiliation{Data Science and Learning Division, Argonne National Laboratory, Argonne, Illinois 60439, USA}

\author{Timothy R. Prisk}
\affiliation{Center for Neutron Research, National Institute of Standards and Technology, Gaithersburg, MD 20899-6100, USA}
\affiliation{Division of Chemistry and Chemical Engineering, California Institute of Technology, Pasadena, California 91125, USA}

\author{Garfield Warren}
\affiliation{Department of Physics, Indiana University, Bloomington, IN 47408, USA}

\author{Paul E. Sokol}
\affiliation{Department of Physics, Indiana University, Bloomington, IN 47408, USA}

\begin{abstract}
\end{abstract}

\maketitle
    The realization of experimental platforms exhibiting one dimensional (1D) quantum phenomena has been elusive, due to their inherent lack of stability, with a few notable exceptions including spin chains \cite{Lake:2005ho}, carbon nanotubes \cite{Bockrath:1999ww} and ultracold low-density gasses \cite{Paredes:2004fp}. The difficulty of such systems in exhibiting long range order is integral to their effective description in terms of the Tomonaga-Luttinger liquid theory \cite{Tomonaga:1950ak,Luttinger:1963gd,Mattis:1965iq,Haldane:1981eh}.  Recently, it has been proposed that the bosonic superfluid $^4$He could realize a 1D quantum system \cite{DelMaestro:2011dh} beyond the Luttinger liquid paradigm \cite{Bertaina:2016gu}.  Here we describe an experimental observation of this behavior using nanoengineering by preplating a porous material with a noble gas to enhance dimensional reduction. The resulting excitations of the confined $^4$He are qualitatively different than 3D and 2D superfluid helium, and can be analyzed in terms of a mobile impurity in a Luttinger liquid \cite{Imambekov:2012ho} allowing for the characterization of the emergent quantum liquid. The confined helium system offers the possibility of tuning via pressure--- from weakly interacting, all the way to the super Tonks-Girardeau gas of strongly interacting hard-core particles. 

\begin{figure}[t!]
    \centering
    \includegraphics[width=\columnwidth]{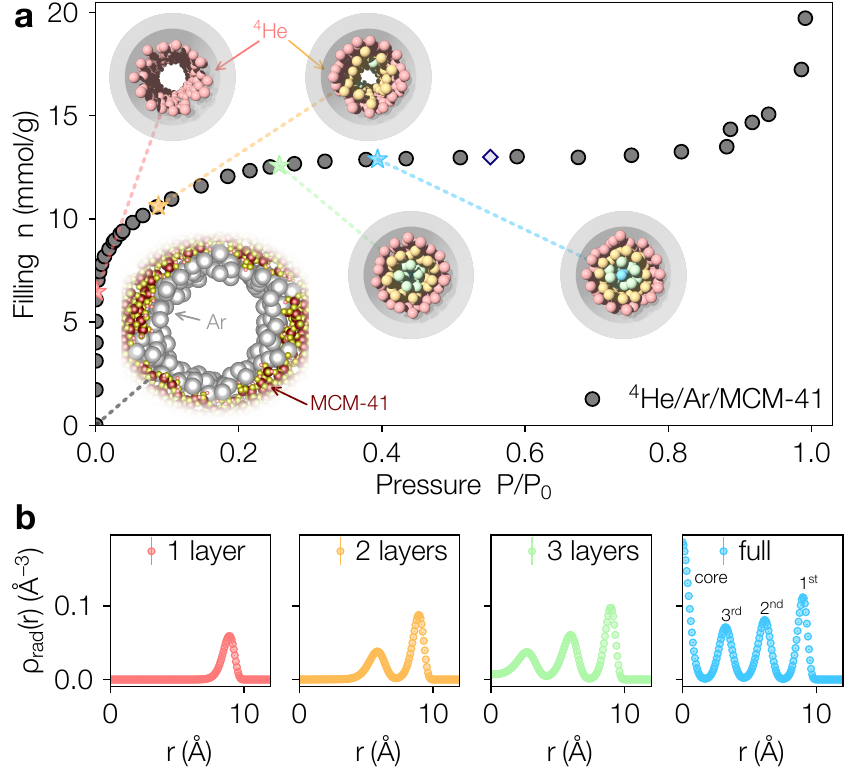}
    \caption{\textbf{Adsorption and structure inside nanopores.}
    \textbf{a} Dark grey circles illustrate the adsorption behavior of $^4$He at \SI{4.2}{\kelvin} into MCM-41 pre-plated with a monolayer of Ar gas as the pressure is increased.  Here $P_0$ is the bulk equilibrium vapor pressure of $^4$He. The colored stars indicate the fillings where completion of $^4$He layers occurs with the call-out inset images showing quantum Monte Carlo configurations of a cross-section of MCM-41 with an equilibrated Ar layer (light grey spheres) at $P/P_0 = 0$, and the developing layers of $^4$He (1 layer to 3 layers plus central core) as the pressure is increased.  Here the Ar is represented as a cylindrical shell for clarity. The light purple diamond indicates the filling at which experimental inelastic neutron scattering measurements were performed at $Q_{\rm in} = \SI{4.0}{\angstrom^{-1}}$ corresponding to completely filled pores.  \textbf{b} Quantum Monte Carlo results for the radial number density of atoms $\rho_{\rm rad}(r)$ inside nanopores at $T = \SI{1.6}{\kelvin}$ where the scattering experiments were performed. Colors correspond to the starred filling fractions in panel \textbf{a}. As the pressure is increased, the $^4$He atoms form a series of concentric layers, with the density of the outer layers also increasing. }
\label{fig:4Heisotherm}
\end{figure}

\begin{figure*}[t]
    \centering
\includegraphics{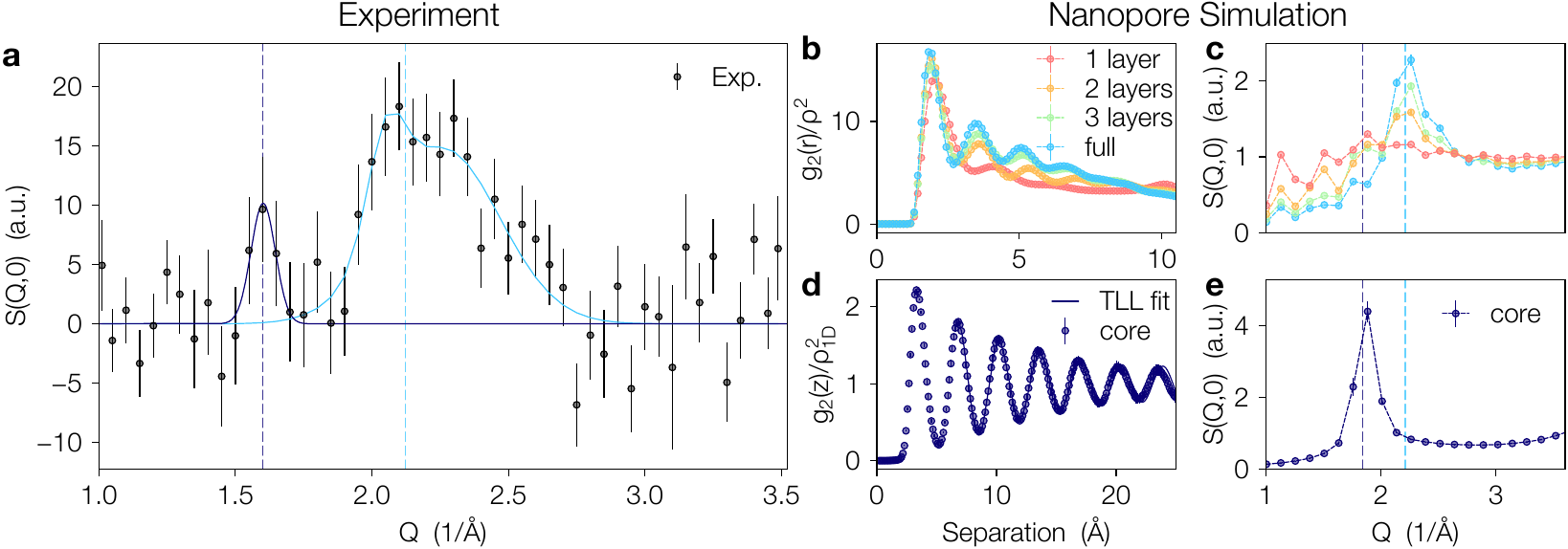}
\caption{%
\textbf{Elastic scattering from $^4$He confined inside MCM-41.} \textbf{a} Experimental $S(Q,0)$ from helium confined to Ar pre-plated pores with an incident wavevector $Q_{\rm in}$ of \SI{2.5}{\angstrom^{-1}} at $T=\SI{1.6}{\kelvin}$.  The presented data is for the completely filled pore with (\SI{13}{\milli\mol\per\gram}) with the scattering from the Ar and boundary layer helium (\SI{8.68}{\milli\mol\per\gram}) subtracted (see Supplementary information for the raw data). The shown uncertainties are the statistical errors after the subtraction. The light blue line is a fit of two Gaussians to the scattering centered around \SI{2.1}{\angstrom^{-1}} originating from the second and third layers of helium in the pores.  The purple line is a fit to a Gaussian centered at \SI{1.6}{\angstrom^{-1}} which we attribute to helium at the pore center. Panels \textbf{b} -- \textbf{e} show the results of quantum Monte Carlo simulations for the structure of $^4$He confined inside a smooth Ar pre-plated nanopore at $T = \SI{1.6}{\kelvin}$.  \textbf{b} shows the radially averaged density-density correlation function $g_2(r) = \expval{\rho(r)\rho(0)}$ in units of the density $\rho = N/(\pi R^2 L)$ with $N$ the total number of particles inside a pore of length $L$ and radius $R$ for the four filling fractions highlighted in Fig.~\ref{fig:4Heisotherm}. \textbf{c} The resulting static structure factor for all atoms in the pore.  \textbf{d} The projected density-density correlations for separations measured along the pore for only those atoms in the central core of the fully filled pore along with a fit to the low energy prediction from Tomonaga-Luttinger liquid theory (solid line). The resulting core-only structure factor in \textbf{e} shows a peak at $2k_{\rm F} = 2\pi \rho_{1D}$(indicated by a vertical dashed line) where $\rho_{1D} = N/L$. The lighter blue vertical line (also in panel \textbf{a}) corresponds to the separation which minimizes the bulk $^4$He -- $^4$He interaction potential that controls the structure in the strongly adsorbed outer layers.
}
\label{fig:HR_confined_elastic_ll}
\end{figure*}

\begin{figure}[h!]
    \centering
    \includegraphics[width=\columnwidth]{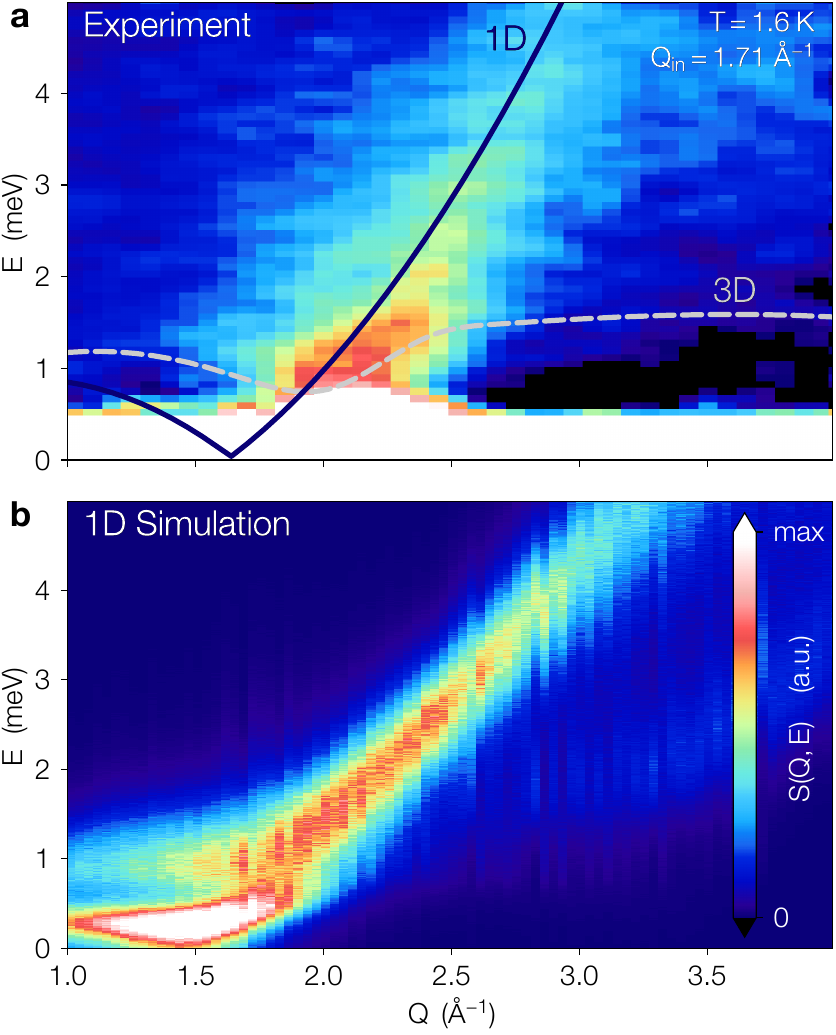}
    \caption{\textbf{Inelastic scattering of a 1D quantum liquid}. \textbf{a} The dynamic structure factor, $S(Q,E)$, from helium confined in the Ar plated pores of MCM-41.  The scattering was measured at a filling of \SI{13}{\milli\mol\per\gram}, corresponding to full pores, and the background from the Ar/MCM-41 matrix has been removed with the elastic scattering suppressed.  The most prominent feature is the strong excitation that begins at $\sim \SI{1.6}{\angstrom^{-1}}$ at $E=0$ and extends to high energy.  This is quite distinct from the scattering that would be expected from either bulk $^4$He or that of  $^4$He confined in larger pores where the scattering would be described by the phonon-maxon-roton excitation curve (grey dashed line). For the bulk liquid, the scattering would be most intense around the roton minimum at $Q_{r} \simeq \SI{1.9}{\angstrom^{-1}}$ and would plateau at twice the roton energy ($\sim \SI{1.5}{\milli\eV}$).  The purple line shows a theoretical prediction for the threshold energy of a purely 1D quantum liquid of hard spheres with emergent Fermi wavevector $2k_{\rm F}=\SI{1.6}{\angstrom^{-1}}$ and dimensionless Tomonaga-Luttinger liquid parameter $K \simeq 1.2$.  The value for $k_{\rm F}$ has been obtained from $S(Q,0)$, while $K$ comes from fitting the inelastic branch of $S(Q,E)$ (see Fig.~\ref{fig:Peak_fit}).  \textbf{b} The dynamic structure factor of a purely 1D system of $^4$He at $T=\SI{1.6}{\kelvin}$ obtained via numerical analytic continuation of the imaginary time scattering function measured via quantum Monte Carlo for a system with $L=\SI{200}{\angstrom}$. The density was chosen to produce a value of $K$ similar to that seen in the experiment.  
\label{fig:INS}
}
\end{figure}

The helium isotopes $^3$He and $^4$He have long served as model systems for precision tests of theories of strongly interacting quantum matter and phase transitions for bosons, fermions, and mixed statistics systems.  This has been spectacularly successful in two \cite{Bishop:1978iq} and three dimensions \cite{Godfrin:2021fj}, where the dynamics can be understood in terms of quasi-particles which retain the original nature of the system through renormalized properties.  In one spatial dimension (1D), the fundamental excitations of a quantum liquid of helium will be collective in nature, and at long wavelengths and low energies, it should be described via the linear hydrodynamics of Tomonaga-Luttinger liquid (TLL) theory. Within this picture, the distinction between bosons and fermions begins to break down and access to both bosonic and fermionic isotopes makes 1D helium an exciting system to explore.  This has motivated a number of experimental \cite{Duc:2015pd,Botimer:2016pl,Wada:2001jb,Prisk:2013cu,Bossy:2019qd,Vekhov:2012hg} and theoretical \cite{Boninsegni:2007qe,DelMaestro:2011dh,Markic:2015bu,Nichols:2020of} studies in quasi-1D confinement, providing tantalizing evidence of low-dimensional or TLL behavior.  However, as helium is a neutral quantum liquid, the route to 1D requires physical confinement in two out of three dimensions at the level of a single nanometer -- the scale of the superfluid coherence length at low temperature -- a difficult feat for real devices. In this paper, we introduce a nanoengineered confining environment, and report elastic and inelastic neutron scattering measurements that demonstrate the existence of one dimensional bosonic $^4$He.

\begin{figure*}[ht]
    \centering
    \includegraphics{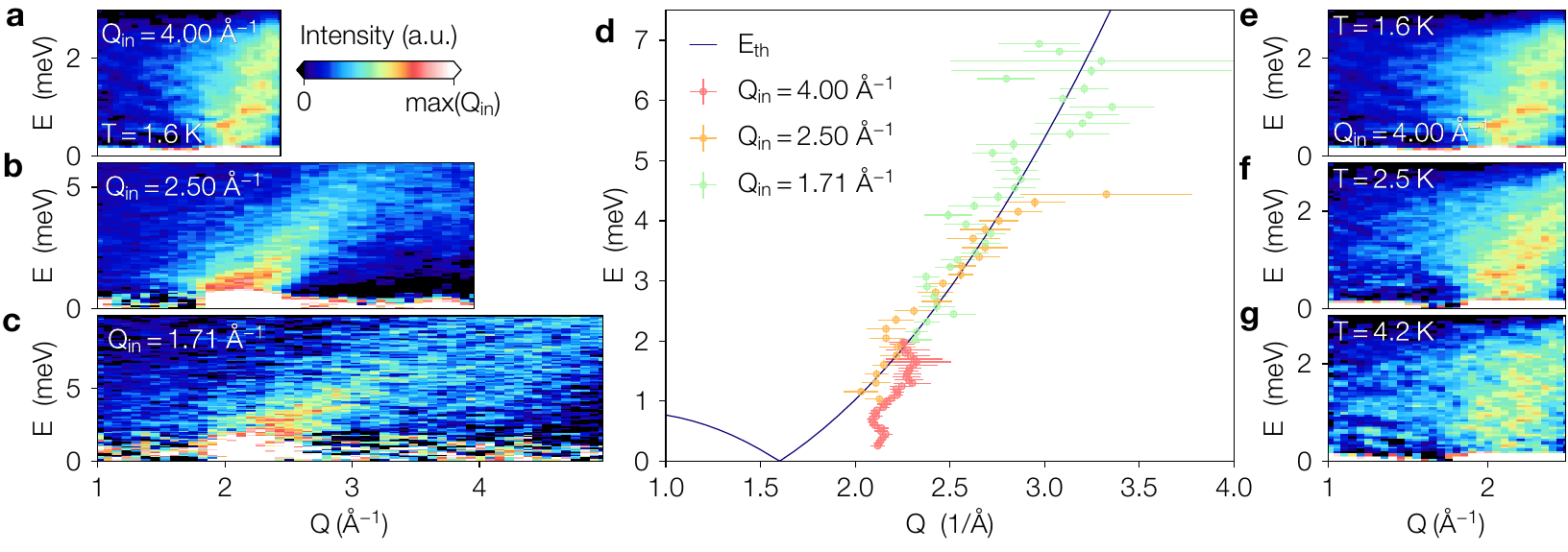}
    \caption{\textbf{Energy and temperature dependence of the inelastic scattering branch.} Left panels: Inelastic scattering from confined $^4$He at various incident wave vectors covering different $Q - E$ ranges with different resolutions.  The plots correspond to incident wavevectors of: \textbf{a} \SI{4.00}{\angstrom^{-1}}, \textbf{b} \SI{2.50}{\angstrom^{-1}}, and  \textbf{c} \SI{1.71}{\angstrom^{-1}}. The $Q$ scales are the same for all three values, but the energy ranges are different.  \textbf{d} the peak excitation energies obtained by fitting a Gaussian in energy to the inelastic scattering in panels \textbf{a} -- \textbf{c}. The solid line is the best fit to the threshold energy of a purely 1D quantum liquid of hard spheres. The fit was obtained by fixing $k_{\rm F}=\SI{0.8}{\angstrom^{-1}}$, as extracted from the $S(Q,0)$ measurement, and yielded $K=1.18^{+0.38}_{-0.20}$. Right Panels: Inelastic scattering from confined helium at various temperatures above and below the bulk superfluid transition.  The measurements have fixed $Q_{
    \rm in}= \SI{4.00}{\angstrom^{-1}}$ and temperatures of: \textbf{e} \SI{1.6}{\kelvin}, \textbf{f} \SI{2.5}{\kelvin}, and \textbf{g} \SI{4.2}{\kelvin}.  The scattering intensity decreases slightly with increasing temperature but remains large even at the highest temperature studied.}
    \label{fig:Peak_fit}
\end{figure*}


Our confinement platform consists of $^4$He adsorbed inside MCM-41, a mesoporous material with a hierarchical structure consisting of cylindrical pores arranged in a hexagonal lattice, pre-plated with a single monolayer of argon \cite{Nichols:2020of}.  The MCM-41 sample had an as-synthesized pore diameter of $\SI{3.0 \pm 0.3}{\nano\meter}$ as determined by N$_2$ isotherms, known to be too large to yield 1D confinement \cite{Bryan:2017hb}. The effective pore diameter was reduced to \SI{2}{\nano\meter} through a pre-treatment step where Ar gas was added at \SI{90}{\kelvin}. The resulting Ar/MCM-41 confining environment was characterized via a $^4$He adsorption isotherm at $T=\SI{4.2}{\kelvin}$ combined with molecular dynamics and quantum Monte Carlo simulations as shown in Figure~\ref{fig:4Heisotherm} (for sample preparation and simulation details see the Methods section).  The results demonstrate that the helium atoms are initially strongly bound to the Ar-plated pore walls at low fillings, and completely fill the pores at \SI{13}{\milli\mol\per\gram}. At intermediate filling, nested cylindrical layers of helium are formed and particle exchanges between them are suppressed.

Understanding this microscopic structure will be crucial to interpret excitations inside the pores as measured by both elastic and inelastic neutron scattering. It is important to note that the elastic scattering with zero energy transfer $E=0$ at wavevector $Q$, $S(Q,0)$, is distinct from the static structure factor $S(Q)= \int_{-\infty}^{\infty} S(Q,E) \,dE$ which averages the collective motion of the system over all time scales.  The elastic scattering, in contrast, probes the static behavior of the system. For example, while bulk liquid helium exhibits a well defined $S(Q)$ reflecting the dynamic correlations in the liquid \cite{Godfrin:2021fj}, the elastic scattering is identically zero. Helium confined in various porous media \cite{Bossy:2019qd} does exhibit elastic scattering due to the presence of solid layers strongly adsorbed on the pore boundaries (see Fig.~\ref{fig:4Heisotherm}\textbf{b}). However, in previously studied pores with large radii that are not yet in the quasi-1D regime, dense liquid layers and a bulk-like liquid in the center of the pore do not contribute to the elastic scattering.

Experimental results for elastic scattering from helium confined inside Ar pre-plated MCM-41 are shown in Figure~\ref{fig:HR_confined_elastic_ll}\textbf{a}, where two clear features are apparent.  (1) There is a broad peak at \SI{\sim 2.1}{\angstrom^{-1}} which we attribute to the second and third strongly adsorbed layers of helium in the pores. We note that due to the existence of the adsorption potential, these layers are more dense than helium in the bulk, and their inter-atomic spacing is consistent with this peak as observed in quantum Monte Carlo simulations (panels \textbf{b} -- \textbf{c}) which show the predicted pair correlation function $g_2(r) = \expval{\rho(r)\rho(0)}$ where $\rho(r)$ is the density of $^4$He and resulting elastic scattering $S(Q,0)$ (see Methods for details).  Here, different curves correspond to the different filling fractions presented in Fig.~\ref{fig:4Heisotherm} and the qualitative agreement between the simulation and experiments supports this picture.  (2) A narrow feature at \SI{\sim 1.6}{\angstrom^{-1}} that we attribute to the atoms at the center of the pore--- the core liquid. This peak has been fit to a Gaussian centered at \SI{1.60 \pm 0.02}{\angstrom^{-1}} which corresponds to an atomic spacing of \SI{3.92 \pm 0.05}{\angstrom}. This is in agreement with simulation results for only those atoms in the 1D central core shown in panels \textbf{d} -- \textbf{e} where weakly decaying oscillations in the density-density correlations along the pore produce a strong peak in the elastic scattering.  Such a peak is predicted by the TLL theory to occur at a wavevector of $2k_{\rm F} = 2\pi\rho_{1D}$ due to the existence of algebraically decaying Bragg peaks in the structure factor \cite{Luther:1974fn} and has been previously observed in numerical simulations of strictly one dimensional $^4$He \cite{Motta:2016ge}.  The emergence of a \emph{Fermi} wavevector is a consequence of 1D where the hard core of the $^4$He -- $^4$He interaction potential dominates making the bosonic system exhibit properties of an ideal gas of spinless fermions. This is consistent with the interpretation that in interacting gapless quantum systems $k_{\rm F}$ corresponds to the smallest momentum at which energy can be absorbed.

Thus we can separate contributions of adsorbed layers and central core atoms, which are highlighted by two vertical dashed lines in panels \textbf{c} and \textbf{e}, and repeated in panel \textbf{a}, supporting the interpretation of the experimental results as demonstrating the existence of
a one dimensional quasi-liquid of helium with a linear density near $\rho_{1D} \approx \SI{0.25}{\angstrom^{-1}}$.  Further evidence comes from the relative intensity of this peak and that due to the first and second layers, which is in agreement with the theoretical number of atoms in the core liquid versus shell region (see Supplementary information). 

We now turn to probing the dynamics of the confined helium within the nanopores. 
Figures~\ref{fig:INS}\textbf{a} and \ref{fig:Peak_fit} show the measured inelastic scattering as captured by the dynamic structure factor $S(Q,E)$ while Figure~\ref{fig:INS}\textbf{b} shows a quantum Monte Carlo prediction for a purely 1D model of $^4$He using differential evolution to analytically continue imaginary time correlations to real frequencies \cite{Nichols:2022d}.  As described above, the observed $S(Q,E)$ arises primarily for the core quasi-liquid, with the immobile solid layers contributing very little to the inelastic scattering.  Figure 3\textbf{a} demonstrates a single well defined inelastic feature which begins at \SI{\sim 1.6}{\angstrom^{-1}} and its energy increases monotonically and smoothly with increasing wavevector $Q$. We note that this is qualitatively distinct from bulk liquid helium (dashed line) which exhibits a well defined excitation spectrum with a linear phonon mode at low $Q$ ($E_{\rm phonon} \propto Q$) and a very intense roton mode with gap $\Delta$ at intermediate $Q = Q_r \approx \SI{1.9}{\angstrom^{-1}}$  ($E_{\rm roton} \approx \Delta + C(Q-Q_r)^2$) and a plateau at \SI{\sim 1.5}{meV} at $Q > Q_r$ due to two roton processes. 

While at low-$Q$, the harmonic Tomonaga-Luttinger liquid theory describes a phonon branch where the only excitations are density waves propagating with velocity $v$ such that $S(Q,E) \simeq Q \delta(E-\hbar v Q)$ \cite{Dzyaloshinskii:1974un},  at higher $Q$, the dynamic structure factor is known to be markedly different.  Here, the effects of band curvature introduce edge singularities in $S(Q,E)$ \cite{Pustilnik:2006bs,Pereira:2008bw, Imambekov:2012ho} where low-energy excitations can only proliferate above an energy threshold $E_{\rm th}$. By interpreting the excitations of $^4$He confined in Ar-preplated MCM-41 in terms of an effective quantum impurity model for a hole propagating in an otherwise linear TLL \cite{Pustilnik:2006bs,Pereira:2008bw,Imambekov:2009ly,Imambekov:2012ho}, the dynamical structure factor can be shown to develop a $Q$-dependent power law singularity at low energies: 
\begin{equation}
S(Q,E) \propto \Theta(E-E_{\rm th}(Q))|E-E_{\rm th}(Q)|^{-\mu(Q)}
\label{eq:SQELL}
\end{equation}
where for a model of hard core bosons \cite{Motta:2016ge,Bertaina:2016gu} 
\begin{equation}
    E_{\rm th}(Q) \simeq \frac{4E_{\rm F}}{K} \qty[ \frac{Q}{2k_{\rm F}} - \qty(\frac{Q}{2k_{\rm F}})^2]
    \label{eq:Ethreshold}
\end{equation}
with $K$ the Luttinger parameter and the appearance of the Fermi energy $E_{\rm F} = \hbar^2 k_{\rm F}^2/(2m)$ further supporting the emergent fermionization of the 1D $^4$He. Here the exponent $\mu(Q)$ is both non-universal (depending on the details of the microscopic interactions and cutoff) as well as momentum dependent, and can vary significantly near $2k_{\rm F}$, even changing sign \cite{Bertaina:2016gu,Motta:2016ge}! We interpret the observed inelastic branch in Figure~\ref{fig:INS} as corresponding to $E_{\rm th}$ and fit the maxima to Eq.~\eqref{eq:Ethreshold} over a broad range of $Q$ (see Figure~\ref{fig:Peak_fit}\textbf{a} -- \textbf{d}) with the value of $2k_{\rm F}$ fixed at \SI{1.6}{\angstrom^{-1}} as determined by the elastic scattering.  This allows us to extract a best fit value of the Luttinger parameter of $K =1.18^{+0.38}_{-0.20}$ which is consistent with microscopic predictions for $^4$He inside smooth nanopores from quantum Monte Carlo \cite{DelMaestro:2011dh} as well as a model of hardcore bosons where $K = (1-\rho_{1D} a_{1D})^2$ with $a_{1D}$ the effective 1D scattering length \cite{Olshanii:1998cf}.

The emergence of a 1D quasi-liquid is further supported by Figure~\ref{fig:Peak_fit}\textbf{e} -- \textbf{g} which shows the weak temperature dependence of the inelastic scattering below $T = \SI{4.2}{\kelvin}$ where the main dispersing feature becomes more diffuse as the temperature is increased. This is consistent with the extracted value of $K \simeq 1.2$ which sets the relevant scale below which we expect to observe TLL behavior to be $T \ll \hbar v Q/k_{\rm B} \simeq 4E_{\rm F}/(K k_{\rm B}) \simeq \SI{13}{\kelvin}$ for $Q \simeq 2k_{\rm F}$ where $k_{\rm B}$ is Boltzmann's constant. Thus the helium core liquid still retains its 1D excitation spectrum even above the bulk superfluid temperature. 

We have created a novel nano-engineered confinement environment for helium that has allowed us to observe 1D quantum liquid behavior beyond the Luttinger liquid paradigm and obtained information on the microscopic structure inside the pores via quantum Monte Carlo simulations. 
The next steps are many, including exploring the ability to control the filling fraction via pressure near the full pore regime to tune the density (and thus Luttinger parameter) of the confined liquid, manifest as a modified slope of the inelastic threshold $E_{\rm th}$ in the excitation spectrum. Replacing bosonic $^4$He with the fermionic isotope $^3$He is even more exciting and would potentially open the door for the observation of spin-mass separation, where spin and density waves propagate with different excitation velocities.


\FloatBarrier

\nocite{apsrev42Control}
\bibliographystyle{apsrev4-2}
\bibliography{refs}

\begin{thebibliography}{44}%
\makeatletter
\providecommand \@ifxundefined [1]{%
 \@ifx{#1\undefined}
}%
\providecommand \@ifnum [1]{%
 \ifnum #1\expandafter \@firstoftwo
 \else \expandafter \@secondoftwo
 \fi
}%
\providecommand \@ifx [1]{%
 \ifx #1\expandafter \@firstoftwo
 \else \expandafter \@secondoftwo
 \fi
}%
\providecommand \natexlab [1]{#1}%
\providecommand \enquote  [1]{``#1''}%
\providecommand \bibnamefont  [1]{#1}%
\providecommand \bibfnamefont [1]{#1}%
\providecommand \citenamefont [1]{#1}%
\providecommand \href@noop [0]{\@secondoftwo}%
\providecommand \href [0]{\begingroup \@sanitize@url \@href}%
\providecommand \@href[1]{\@@startlink{#1}\@@href}%
\providecommand \@@href[1]{\endgroup#1\@@endlink}%
\providecommand \@sanitize@url [0]{\catcode `\\12\catcode `\$12\catcode
  `\&12\catcode `\#12\catcode `\^12\catcode `\_12\catcode `\%12\relax}%
\providecommand \@@startlink[1]{}%
\providecommand \@@endlink[0]{}%
\providecommand \url  [0]{\begingroup\@sanitize@url \@url }%
\providecommand \@url [1]{\endgroup\@href {#1}{\urlprefix }}%
\providecommand \urlprefix  [0]{URL }%
\providecommand \Eprint [0]{\href }%
\providecommand \doibase [0]{https://doi.org/}%
\providecommand \selectlanguage [0]{\@gobble}%
\providecommand \bibinfo  [0]{\@secondoftwo}%
\providecommand \bibfield  [0]{\@secondoftwo}%
\providecommand \translation [1]{[#1]}%
\providecommand \BibitemOpen [0]{}%
\providecommand \bibitemStop [0]{}%
\providecommand \bibitemNoStop [0]{.\EOS\space}%
\providecommand \EOS [0]{\spacefactor3000\relax}%
\providecommand \BibitemShut  [1]{\csname bibitem#1\endcsname}%
\let\auto@bib@innerbib\@empty
\bibitem [{\citenamefont {Lake}\ \emph {et~al.}(2005)\citenamefont {Lake},
  \citenamefont {Tennant}, \citenamefont {Frost},\ and\ \citenamefont
  {Nagler}}]{Lake:2005ho}%
  \BibitemOpen
  \bibfield  {author} {\bibinfo {author} {\bibfnamefont {B.}~\bibnamefont
  {Lake}}, \bibinfo {author} {\bibfnamefont {D.~A.}\ \bibnamefont {Tennant}},
  \bibinfo {author} {\bibfnamefont {C.~D.}\ \bibnamefont {Frost}},\ and\
  \bibinfo {author} {\bibfnamefont {S.~E.}\ \bibnamefont {Nagler}},\ }\bibfield
   {title} {\bibinfo {title} {{Q}uantum criticality and universal scaling of a
  quantum antiferromagnet},\ }\href {https://doi.org/10.1038/nmat1327}
  {\bibfield  {journal} {\bibinfo  {journal} {Nature Mat.}\ }\textbf {\bibinfo
  {volume} {4}},\ \bibinfo {pages} {329} (\bibinfo {year} {2005})}\BibitemShut
  {NoStop}%
\bibitem [{\citenamefont {Bockrath}\ \emph {et~al.}(1999)\citenamefont
  {Bockrath}, \citenamefont {Cobden}, \citenamefont {Lu}, \citenamefont
  {Rinzler}, \citenamefont {Smalley}, \citenamefont {Balents},\ and\
  \citenamefont {McEuen}}]{Bockrath:1999ww}%
  \BibitemOpen
  \bibfield  {author} {\bibinfo {author} {\bibfnamefont {M.}~\bibnamefont
  {Bockrath}}, \bibinfo {author} {\bibfnamefont {D.}~\bibnamefont {Cobden}},
  \bibinfo {author} {\bibfnamefont {J.}~\bibnamefont {Lu}}, \bibinfo {author}
  {\bibfnamefont {A.}~\bibnamefont {Rinzler}}, \bibinfo {author} {\bibfnamefont
  {R.}~\bibnamefont {Smalley}}, \bibinfo {author} {\bibfnamefont
  {L.}~\bibnamefont {Balents}},\ and\ \bibinfo {author} {\bibfnamefont
  {P.}~\bibnamefont {McEuen}},\ }\bibfield  {title} {\bibinfo {title}
  {{Luttinger-liquid behaviour in carbon nanotubes}},\ }\href
  {http://www.nature.com/nature/journal/v397/n6720/full/397598a0.html?free=2}
  {\bibfield  {journal} {\bibinfo  {journal} {Nature}\ }\textbf {\bibinfo
  {volume} {397}},\ \bibinfo {pages} {598} (\bibinfo {year}
  {1999})}\BibitemShut {NoStop}%
\bibitem [{\citenamefont {Paredes}\ \emph {et~al.}(2004)\citenamefont
  {Paredes}, \citenamefont {Widera}, \citenamefont {Murg}, \citenamefont
  {Mandel}, \citenamefont {F{\"o}lling}, \citenamefont {Cirac}, \citenamefont
  {Shlyapnikov}, \citenamefont {H{\"a}nsch},\ and\ \citenamefont
  {Bloch}}]{Paredes:2004fp}%
  \BibitemOpen
  \bibfield  {author} {\bibinfo {author} {\bibfnamefont {B.}~\bibnamefont
  {Paredes}}, \bibinfo {author} {\bibfnamefont {A.}~\bibnamefont {Widera}},
  \bibinfo {author} {\bibfnamefont {V.}~\bibnamefont {Murg}}, \bibinfo {author}
  {\bibfnamefont {O.}~\bibnamefont {Mandel}}, \bibinfo {author} {\bibfnamefont
  {S.}~\bibnamefont {F{\"o}lling}}, \bibinfo {author} {\bibfnamefont
  {I.}~\bibnamefont {Cirac}}, \bibinfo {author} {\bibfnamefont {G.~V.}\
  \bibnamefont {Shlyapnikov}}, \bibinfo {author} {\bibfnamefont {T.~W.}\
  \bibnamefont {H{\"a}nsch}},\ and\ \bibinfo {author} {\bibfnamefont
  {I.}~\bibnamefont {Bloch}},\ }\bibfield  {title} {\bibinfo {title}
  {{Tonks{\textendash}Girardeau gas of ultracold atoms in an optical
  lattice}},\ }\href {http://www.nature.com/doifinder/10.1038/nature02530}
  {\bibfield  {journal} {\bibinfo  {journal} {Nature}\ }\textbf {\bibinfo
  {volume} {429}},\ \bibinfo {pages} {277} (\bibinfo {year}
  {2004})}\BibitemShut {NoStop}%
\bibitem [{\citenamefont {Tomonaga}(1950)}]{Tomonaga:1950ak}%
  \BibitemOpen
  \bibfield  {author} {\bibinfo {author} {\bibfnamefont {S.~I.}\ \bibnamefont
  {Tomonaga}},\ }\bibfield  {title} {\bibinfo {title} {{R}emarks on {B}loch's
  {M}ethod of {S}ound {W}aves applied to {M}any-{F}ermion {P}roblems},\ }\href
  {https://doi.org/10.1143/ptp/5.4.544} {\bibfield  {journal} {\bibinfo
  {journal} {Prog. Theor. Phys.}\ }\textbf {\bibinfo {volume} {5}},\ \bibinfo
  {pages} {544} (\bibinfo {year} {1950})}\BibitemShut {NoStop}%
\bibitem [{\citenamefont {Luttinger}(1963)}]{Luttinger:1963gd}%
  \BibitemOpen
  \bibfield  {author} {\bibinfo {author} {\bibfnamefont {J.~M.}\ \bibnamefont
  {Luttinger}},\ }\bibfield  {title} {\bibinfo {title} {{A}n {E}xactly
  {S}oluble {M}odel of a {M}any-{F}ermion {S}ystem},\ }\href
  {https://doi.org/10.1063/1.1704046} {\bibfield  {journal} {\bibinfo
  {journal} {J. Math. Phys.}\ }\textbf {\bibinfo {volume} {4}},\ \bibinfo
  {pages} {1154} (\bibinfo {year} {1963})}\BibitemShut {NoStop}%
\bibitem [{\citenamefont {Mattis}\ and\ \citenamefont
  {Lieb}(1965)}]{Mattis:1965iq}%
  \BibitemOpen
  \bibfield  {author} {\bibinfo {author} {\bibfnamefont {D.~C.}\ \bibnamefont
  {Mattis}}\ and\ \bibinfo {author} {\bibfnamefont {E.~H.}\ \bibnamefont
  {Lieb}},\ }\bibfield  {title} {\bibinfo {title} {{E}xact {S}olution of a
  {M}any-{F}ermion {S}ystem and {I}ts {A}ssociated {B}oson {F}ield},\ }\href
  {https://doi.org/10.1063/1.1704281} {\bibfield  {journal} {\bibinfo
  {journal} {J. Math. Phys.}\ }\textbf {\bibinfo {volume} {6}},\ \bibinfo
  {pages} {304} (\bibinfo {year} {1965})}\BibitemShut {NoStop}%
\bibitem [{\citenamefont {Haldane}(1981)}]{Haldane:1981eh}%
  \BibitemOpen
  \bibfield  {author} {\bibinfo {author} {\bibfnamefont {F.~D.~M.}\
  \bibnamefont {Haldane}},\ }\bibfield  {title} {\bibinfo {title} {{Effective
  Harmonic-Fluid Approach to Low-Energy Properties of One-Dimensional Quantum
  Fluids}},\ }\href {https://doi.org/10.1103/PhysRevLett.47.1840} {\bibfield
  {journal} {\bibinfo  {journal} {Phys. Rev. Lett.}\ }\textbf {\bibinfo
  {volume} {47}},\ \bibinfo {pages} {1840} (\bibinfo {year}
  {1981})}\BibitemShut {NoStop}%
\bibitem [{\citenamefont {Del~Maestro}\ \emph {et~al.}(2011)\citenamefont
  {Del~Maestro}, \citenamefont {Boninsegni},\ and\ \citenamefont
  {Affleck}}]{DelMaestro:2011dh}%
  \BibitemOpen
  \bibfield  {author} {\bibinfo {author} {\bibfnamefont {A.}~\bibnamefont
  {Del~Maestro}}, \bibinfo {author} {\bibfnamefont {M.}~\bibnamefont
  {Boninsegni}},\ and\ \bibinfo {author} {\bibfnamefont {I.}~\bibnamefont
  {Affleck}},\ }\bibfield  {title} {\bibinfo {title} {{$^{4}$He Luttinger
  Liquid in Nanopores}},\ }\href
  {http://link.aps.org/doi/10.1103/PhysRevLett.106.105303} {\bibfield
  {journal} {\bibinfo  {journal} {Phys. Rev. Lett.}\ }\textbf {\bibinfo
  {volume} {106}},\ \bibinfo {pages} {105303} (\bibinfo {year}
  {2011})}\BibitemShut {NoStop}%
\bibitem [{\citenamefont {Bertaina}\ \emph {et~al.}(2016)\citenamefont
  {Bertaina}, \citenamefont {Motta}, \citenamefont {Rossi}, \citenamefont
  {Vitali},\ and\ \citenamefont {Galli}}]{Bertaina:2016gu}%
  \BibitemOpen
  \bibfield  {author} {\bibinfo {author} {\bibfnamefont {G.}~\bibnamefont
  {Bertaina}}, \bibinfo {author} {\bibfnamefont {M.}~\bibnamefont {Motta}},
  \bibinfo {author} {\bibfnamefont {M.}~\bibnamefont {Rossi}}, \bibinfo
  {author} {\bibfnamefont {E.}~\bibnamefont {Vitali}},\ and\ \bibinfo {author}
  {\bibfnamefont {D.~E.}\ \bibnamefont {Galli}},\ }\bibfield  {title} {\bibinfo
  {title} {{One-Dimensional Liquid 4He: Dynamical Properties beyond
  Luttinger-Liquid Theory}},\ }\href
  {http://journals.aps.org/prl/abstract/10.1103/PhysRevLett.116.135302}
  {\bibfield  {journal} {\bibinfo  {journal} {Phys. Rev. Lett.}\ }\textbf
  {\bibinfo {volume} {116}},\ \bibinfo {pages} {135302} (\bibinfo {year}
  {2016})}\BibitemShut {NoStop}%
\bibitem [{\citenamefont {Imambekov}\ \emph {et~al.}(2012)\citenamefont
  {Imambekov}, \citenamefont {Schmidt},\ and\ \citenamefont
  {Glazman}}]{Imambekov:2012ho}%
  \BibitemOpen
  \bibfield  {author} {\bibinfo {author} {\bibfnamefont {A.}~\bibnamefont
  {Imambekov}}, \bibinfo {author} {\bibfnamefont {T.~L.}\ \bibnamefont
  {Schmidt}},\ and\ \bibinfo {author} {\bibfnamefont {L.~I.}\ \bibnamefont
  {Glazman}},\ }\bibfield  {title} {\bibinfo {title} {{One-dimensional quantum
  liquids: Beyond the Luttinger liquid paradigm}},\ }\href
  {https://link.aps.org/doi/10.1103/RevModPhys.84.1253} {\bibfield  {journal}
  {\bibinfo  {journal} {Rev. Mod. Phys.}\ }\textbf {\bibinfo {volume} {84}},\
  \bibinfo {pages} {1253} (\bibinfo {year} {2012})}\BibitemShut {NoStop}%
\bibitem [{\citenamefont {Bishop}\ and\ \citenamefont
  {Reppy}(1978)}]{Bishop:1978iq}%
  \BibitemOpen
  \bibfield  {author} {\bibinfo {author} {\bibfnamefont {D.~J.}\ \bibnamefont
  {Bishop}}\ and\ \bibinfo {author} {\bibfnamefont {J.~D.}\ \bibnamefont
  {Reppy}},\ }\bibfield  {title} {\bibinfo {title} {{S}tudy of the {S}uperfluid
  {T}ransition in {T}wo-{Dimensional{H}e}4{F}ilms},\ }\href
  {https://doi.org/10.1103/physrevlett.40.1727} {\bibfield  {journal} {\bibinfo
   {journal} {Phys. Rev. Lett.}\ }\textbf {\bibinfo {volume} {40}},\ \bibinfo
  {pages} {1727} (\bibinfo {year} {1978})}\BibitemShut {NoStop}%
\bibitem [{\citenamefont {Godfrin}\ \emph {et~al.}(2021)\citenamefont
  {Godfrin}, \citenamefont {Beauvois}, \citenamefont {Sultan}, \citenamefont
  {Krotscheck}, \citenamefont {Dawidowski}, \citenamefont {F{\aa}k},\ and\
  \citenamefont {Ollivier}}]{Godfrin:2021fj}%
  \BibitemOpen
  \bibfield  {author} {\bibinfo {author} {\bibfnamefont {H.}~\bibnamefont
  {Godfrin}}, \bibinfo {author} {\bibfnamefont {K.}~\bibnamefont {Beauvois}},
  \bibinfo {author} {\bibfnamefont {A.}~\bibnamefont {Sultan}}, \bibinfo
  {author} {\bibfnamefont {E.}~\bibnamefont {Krotscheck}}, \bibinfo {author}
  {\bibfnamefont {J.}~\bibnamefont {Dawidowski}}, \bibinfo {author}
  {\bibfnamefont {B.}~\bibnamefont {F{\aa}k}},\ and\ \bibinfo {author}
  {\bibfnamefont {J.}~\bibnamefont {Ollivier}},\ }\bibfield  {title} {\bibinfo
  {title} {{D}ispersion relation of {L}andau elementary excitations and
  thermodynamic properties of superfluid {H}e4},\ }\href
  {https://doi.org/10.1103/physrevb.103.104516} {\bibfield  {journal} {\bibinfo
   {journal} {Phys. Rev. B}\ }\textbf {\bibinfo {volume} {103}},\ \bibinfo
  {pages} {104516} (\bibinfo {year} {2021})}\BibitemShut {NoStop}%
\bibitem [{\citenamefont {Duc}\ \emph {et~al.}(2015)\citenamefont {Duc},
  \citenamefont {Savard}, \citenamefont {Petrescu}, \citenamefont {Rosenow},
  \citenamefont {Maestro},\ and\ \citenamefont {Gervais}}]{Duc:2015pd}%
  \BibitemOpen
  \bibfield  {author} {\bibinfo {author} {\bibfnamefont {P.-F.}\ \bibnamefont
  {Duc}}, \bibinfo {author} {\bibfnamefont {M.}~\bibnamefont {Savard}},
  \bibinfo {author} {\bibfnamefont {M.}~\bibnamefont {Petrescu}}, \bibinfo
  {author} {\bibfnamefont {B.}~\bibnamefont {Rosenow}}, \bibinfo {author}
  {\bibfnamefont {A.~D.}\ \bibnamefont {Maestro}},\ and\ \bibinfo {author}
  {\bibfnamefont {G.}~\bibnamefont {Gervais}},\ }\bibfield  {title} {\bibinfo
  {title} {Critical flow and dissipation in a quasi{\textendash}one-dimensional
  superfluid},\ }\href {https://doi.org/10.1126/sciadv.1400222} {\bibfield
  {journal} {\bibinfo  {journal} {Sci. Adv.}\ }\textbf {\bibinfo {volume}
  {1}},\ \bibinfo {pages} {e1400222} (\bibinfo {year} {2015})}\BibitemShut
  {NoStop}%
\bibitem [{\citenamefont {Botimer}\ and\ \citenamefont
  {Taborek}(2016)}]{Botimer:2016pl}%
  \BibitemOpen
  \bibfield  {author} {\bibinfo {author} {\bibfnamefont {J.}~\bibnamefont
  {Botimer}}\ and\ \bibinfo {author} {\bibfnamefont {P.}~\bibnamefont
  {Taborek}},\ }\bibfield  {title} {\bibinfo {title} {{{P}ressure driven flow
  of superfluid $^4${H}e through a nanopipe}},\ }\href
  {http://dx.doi.org/10.1103/physrevfluids.1.054102} {\bibfield  {journal}
  {\bibinfo  {journal} {Phys. Rev. Fluids}\ }\textbf {\bibinfo {volume} {1}},\
  \bibinfo {pages} {054102} (\bibinfo {year} {2016})}\BibitemShut {NoStop}%
\bibitem [{\citenamefont {Wada}\ \emph {et~al.}(2001)\citenamefont {Wada},
  \citenamefont {Taniguchi}, \citenamefont {Ikegami}, \citenamefont {Inagaki},\
  and\ \citenamefont {Fukushima}}]{Wada:2001jb}%
  \BibitemOpen
  \bibfield  {author} {\bibinfo {author} {\bibfnamefont {N.}~\bibnamefont
  {Wada}}, \bibinfo {author} {\bibfnamefont {J.}~\bibnamefont {Taniguchi}},
  \bibinfo {author} {\bibfnamefont {H.}~\bibnamefont {Ikegami}}, \bibinfo
  {author} {\bibfnamefont {S.}~\bibnamefont {Inagaki}},\ and\ \bibinfo {author}
  {\bibfnamefont {Y.}~\bibnamefont {Fukushima}},\ }\bibfield  {title} {\bibinfo
  {title} {{Helium-4 Bose Fluids Formed in One-Dimensional 18 {\AA} Diameter
  Pores}},\ }\href {http://link.aps.org/doi/10.1103/PhysRevLett.86.4322}
  {\bibfield  {journal} {\bibinfo  {journal} {Phys. Rev. Lett.}\ }\textbf
  {\bibinfo {volume} {86}},\ \bibinfo {pages} {4322} (\bibinfo {year}
  {2001})}\BibitemShut {NoStop}%
\bibitem [{\citenamefont {Prisk}\ \emph {et~al.}(2013)\citenamefont {Prisk},
  \citenamefont {Das}, \citenamefont {Diallo}, \citenamefont {Ehlers},
  \citenamefont {Podlesnyak}, \citenamefont {Wada}, \citenamefont {Inagaki},\
  and\ \citenamefont {Sokol}}]{Prisk:2013cu}%
  \BibitemOpen
  \bibfield  {author} {\bibinfo {author} {\bibfnamefont {T.~R.}\ \bibnamefont
  {Prisk}}, \bibinfo {author} {\bibfnamefont {N.~C.}\ \bibnamefont {Das}},
  \bibinfo {author} {\bibfnamefont {S.~O.}\ \bibnamefont {Diallo}}, \bibinfo
  {author} {\bibfnamefont {G.}~\bibnamefont {Ehlers}}, \bibinfo {author}
  {\bibfnamefont {A.~A.}\ \bibnamefont {Podlesnyak}}, \bibinfo {author}
  {\bibfnamefont {N.}~\bibnamefont {Wada}}, \bibinfo {author} {\bibfnamefont
  {S.}~\bibnamefont {Inagaki}},\ and\ \bibinfo {author} {\bibfnamefont {P.~E.}\
  \bibnamefont {Sokol}},\ }\bibfield  {title} {\bibinfo {title} {{Phases of
  superfluid helium in smooth cylindrical pores}},\ }\href
  {http://link.aps.org/doi/10.1103/PhysRevB.88.014521} {\bibfield  {journal}
  {\bibinfo  {journal} {Phys. Rev. B}\ }\textbf {\bibinfo {volume} {88}},\
  \bibinfo {pages} {014521} (\bibinfo {year} {2013})}\BibitemShut {NoStop}%
\bibitem [{\citenamefont {Bossy}\ \emph {et~al.}(2019)\citenamefont {Bossy},
  \citenamefont {Ollivier},\ and\ \citenamefont {Glyde}}]{Bossy:2019qd}%
  \BibitemOpen
  \bibfield  {author} {\bibinfo {author} {\bibfnamefont {J.}~\bibnamefont
  {Bossy}}, \bibinfo {author} {\bibfnamefont {J.}~\bibnamefont {Ollivier}},\
  and\ \bibinfo {author} {\bibfnamefont {H.~R.}\ \bibnamefont {Glyde}},\
  }\bibfield  {title} {\bibinfo {title} {{{P}honons, rotons, and localized
  {B}ose-{E}instein condensation in liquid $^4${H}e confined in nanoporous
  {F{S}M}-16}},\ }\href
  {https://journals.aps.org/prb/abstract/10.1103/PhysRevB.99.165425} {\bibfield
   {journal} {\bibinfo  {journal} {Phys. Rev. B}\ }\textbf {\bibinfo {volume}
  {99}},\ \bibinfo {pages} {165425} (\bibinfo {year} {2019})}\BibitemShut
  {NoStop}%
\bibitem [{\citenamefont {Vekhov}\ and\ \citenamefont
  {Hallock}(2012)}]{Vekhov:2012hg}%
  \BibitemOpen
  \bibfield  {author} {\bibinfo {author} {\bibfnamefont {Y.}~\bibnamefont
  {Vekhov}}\ and\ \bibinfo {author} {\bibfnamefont {R.~B.}\ \bibnamefont
  {Hallock}},\ }\bibfield  {title} {\bibinfo {title} {{Mass Flux
  Characteristics in Solid $^{4}$He for $T>100$~mK: Evidence for Bosonic
  Luttinger-Liquid Behavior}},\ }\href
  {http://link.aps.org/doi/10.1103/PhysRevLett.109.045303} {\bibfield
  {journal} {\bibinfo  {journal} {Phys. Rev. Lett.}\ }\textbf {\bibinfo
  {volume} {109}},\ \bibinfo {pages} {045303} (\bibinfo {year}
  {2012})}\BibitemShut {NoStop}%
\bibitem [{\citenamefont {Boninsegni}\ \emph {et~al.}(2007)\citenamefont
  {Boninsegni}, \citenamefont {Kuklov}, \citenamefont {Pollet}, \citenamefont
  {Prokof'ev}, \citenamefont {Svistunov},\ and\ \citenamefont
  {Troyer}}]{Boninsegni:2007qe}%
  \BibitemOpen
  \bibfield  {author} {\bibinfo {author} {\bibfnamefont {M.}~\bibnamefont
  {Boninsegni}}, \bibinfo {author} {\bibfnamefont {A.~B.}\ \bibnamefont
  {Kuklov}}, \bibinfo {author} {\bibfnamefont {L.}~\bibnamefont {Pollet}},
  \bibinfo {author} {\bibfnamefont {N.~V.}\ \bibnamefont {Prokof'ev}}, \bibinfo
  {author} {\bibfnamefont {B.~V.}\ \bibnamefont {Svistunov}},\ and\ \bibinfo
  {author} {\bibfnamefont {M.}~\bibnamefont {Troyer}},\ }\bibfield  {title}
  {\bibinfo {title} {{L}uttinger {L}iquid in the {C}ore of a {S}crew
  {D}islocation in {H}elium-4},\ }\href
  {https://doi.org/10.1103/physrevlett.99.035301} {\bibfield  {journal}
  {\bibinfo  {journal} {Phys. Rev. Lett.}\ }\textbf {\bibinfo {volume} {99}},\
  \bibinfo {pages} {035301} (\bibinfo {year} {2007})}\BibitemShut {NoStop}%
\bibitem [{\citenamefont {Marki{\'{c}}}\ and\ \citenamefont
  {Glyde}(2015)}]{Markic:2015bu}%
  \BibitemOpen
  \bibfield  {author} {\bibinfo {author} {\bibfnamefont {L.~V.}\ \bibnamefont
  {Marki{\'{c}}}}\ and\ \bibinfo {author} {\bibfnamefont {H.~R.}\ \bibnamefont
  {Glyde}},\ }\bibfield  {title} {\bibinfo {title} {{S}uperfluidity, {B{E}C},
  and dimensions of {liquid{H}e}4in nanopores},\ }\href
  {https://doi.org/10.1103/physrevb.92.064510} {\bibfield  {journal} {\bibinfo
  {journal} {Phys. Rev. B}\ }\textbf {\bibinfo {volume} {92}},\ \bibinfo
  {pages} {064510} (\bibinfo {year} {2015})}\BibitemShut {NoStop}%
\bibitem [{\citenamefont {Nichols}\ \emph {et~al.}(2020)\citenamefont
  {Nichols}, \citenamefont {Prisk}, \citenamefont {Warren}, \citenamefont
  {Sokol},\ and\ \citenamefont {{Del Maestro}}}]{Nichols:2020of}%
  \BibitemOpen
  \bibfield  {author} {\bibinfo {author} {\bibfnamefont {N.~S.}\ \bibnamefont
  {Nichols}}, \bibinfo {author} {\bibfnamefont {T.~R.}\ \bibnamefont {Prisk}},
  \bibinfo {author} {\bibfnamefont {G.}~\bibnamefont {Warren}}, \bibinfo
  {author} {\bibfnamefont {P.}~\bibnamefont {Sokol}},\ and\ \bibinfo {author}
  {\bibfnamefont {A.}~\bibnamefont {{Del Maestro}}},\ }\bibfield  {title}
  {\bibinfo {title} {{D}imensional reduction of helium-4 inside argon-plated
  {M{C}M}-41 nanopores},\ }\href {https://doi.org/10.1103/physrevb.102.144505}
  {\bibfield  {journal} {\bibinfo  {journal} {Phys. Rev. B}\ }\textbf {\bibinfo
  {volume} {102}},\ \bibinfo {pages} {144505} (\bibinfo {year}
  {2020})}\BibitemShut {NoStop}%
\bibitem [{\citenamefont {Bryan}\ \emph {et~al.}(2017)\citenamefont {Bryan},
  \citenamefont {Prisk}, \citenamefont {Sherline}, \citenamefont {Diallo},\
  and\ \citenamefont {Sokol}}]{Bryan:2017hb}%
  \BibitemOpen
  \bibfield  {author} {\bibinfo {author} {\bibfnamefont {M.~S.}\ \bibnamefont
  {Bryan}}, \bibinfo {author} {\bibfnamefont {T.~R.}\ \bibnamefont {Prisk}},
  \bibinfo {author} {\bibfnamefont {T.~E.}\ \bibnamefont {Sherline}}, \bibinfo
  {author} {\bibfnamefont {S.~O.}\ \bibnamefont {Diallo}},\ and\ \bibinfo
  {author} {\bibfnamefont {P.~E.}\ \bibnamefont {Sokol}},\ }\bibfield  {title}
  {\bibinfo {title} {{Bulklike excitations in nanoconfined liquid helium}},\
  }\href {http://link.aps.org/doi/10.1103/PhysRevB.95.144509} {\bibfield
  {journal} {\bibinfo  {journal} {Phys. Rev. B}\ }\textbf {\bibinfo {volume}
  {95}},\ \bibinfo {pages} {144509} (\bibinfo {year} {2017})}\BibitemShut
  {NoStop}%
\bibitem [{\citenamefont {Luther}\ and\ \citenamefont
  {Peschel}(1974)}]{Luther:1974fn}%
  \BibitemOpen
  \bibfield  {author} {\bibinfo {author} {\bibfnamefont {A.}~\bibnamefont
  {Luther}}\ and\ \bibinfo {author} {\bibfnamefont {I.}~\bibnamefont
  {Peschel}},\ }\bibfield  {title} {\bibinfo {title} {{S}ingle-particle states,
  {K}ohn anomaly, and pairing fluctuations in one dimension},\ }\href
  {https://doi.org/10.1103/physrevb.9.2911} {\bibfield  {journal} {\bibinfo
  {journal} {Phys. Rev. B}\ }\textbf {\bibinfo {volume} {9}},\ \bibinfo {pages}
  {2911} (\bibinfo {year} {1974})}\BibitemShut {NoStop}%
\bibitem [{\citenamefont {Motta}\ \emph {et~al.}(2016)\citenamefont {Motta},
  \citenamefont {Vitali}, \citenamefont {Rossi}, \citenamefont {Galli},\ and\
  \citenamefont {Bertaina}}]{Motta:2016ge}%
  \BibitemOpen
  \bibfield  {author} {\bibinfo {author} {\bibfnamefont {M.}~\bibnamefont
  {Motta}}, \bibinfo {author} {\bibfnamefont {E.}~\bibnamefont {Vitali}},
  \bibinfo {author} {\bibfnamefont {M.}~\bibnamefont {Rossi}}, \bibinfo
  {author} {\bibfnamefont {D.~E.}\ \bibnamefont {Galli}},\ and\ \bibinfo
  {author} {\bibfnamefont {G.}~\bibnamefont {Bertaina}},\ }\bibfield  {title}
  {\bibinfo {title} {{Dynamical structure factor of one-dimensional hard
  rods}},\ }\href {https://link.aps.org/doi/10.1103/PhysRevA.94.043627}
  {\bibfield  {journal} {\bibinfo  {journal} {Phys. Rev. A}\ }\textbf {\bibinfo
  {volume} {94}},\ \bibinfo {pages} {043627} (\bibinfo {year}
  {2016})}\BibitemShut {NoStop}%
\bibitem [{\citenamefont {Nichols}\ \emph {et~al.}(2022)\citenamefont
  {Nichols}, \citenamefont {Sokol},\ and\ \citenamefont {{Del
  Maestro}}}]{Nichols:2022d}%
  \BibitemOpen
  \bibfield  {author} {\bibinfo {author} {\bibfnamefont {N.~S.}\ \bibnamefont
  {Nichols}}, \bibinfo {author} {\bibfnamefont {P.}~\bibnamefont {Sokol}},\
  and\ \bibinfo {author} {\bibfnamefont {A.}~\bibnamefont {{Del Maestro}}},\
  }\bibfield  {title} {\bibinfo {title} {A parameter-free differential
  evolution algorithm for the analytic continuation of imaginary time
  correlation functions},\ }\href@noop {} {\bibfield  {journal} {\bibinfo
  {journal} {arXiv:2201.04155}\ } (\bibinfo {year} {2022})}\BibitemShut
  {NoStop}%
\bibitem [{\citenamefont {Dzyaloshinskii}\ and\ \citenamefont
  {Larkin}(1973)}]{Dzyaloshinskii:1974un}%
  \BibitemOpen
  \bibfield  {author} {\bibinfo {author} {\bibfnamefont {I.~E.}\ \bibnamefont
  {Dzyaloshinskii}}\ and\ \bibinfo {author} {\bibfnamefont {A.~I.}\
  \bibnamefont {Larkin}},\ }\bibfield  {title} {\bibinfo {title} {{Correlation
  functions for a one-dimensional Fermi system with long-range interaction
  (Tomonaga model)}},\ }\href@noop {} {\bibfield  {journal} {\bibinfo
  {journal} {Zh. Eksp. Teor. Fiz.}\ }\textbf {\bibinfo {volume} {65}},\
  \bibinfo {pages} {411} (\bibinfo {year} {1973})},\ \bibinfo {note} {{[Sov.
  Phys. JETP \textbf{38}, 202 (1974)]}}\BibitemShut {NoStop}%
\bibitem [{\citenamefont {Pustilnik}\ \emph {et~al.}(2006)\citenamefont
  {Pustilnik}, \citenamefont {Khodas}, \citenamefont {Kamenev},\ and\
  \citenamefont {Glazman}}]{Pustilnik:2006bs}%
  \BibitemOpen
  \bibfield  {author} {\bibinfo {author} {\bibfnamefont {M.}~\bibnamefont
  {Pustilnik}}, \bibinfo {author} {\bibfnamefont {M.}~\bibnamefont {Khodas}},
  \bibinfo {author} {\bibfnamefont {A.}~\bibnamefont {Kamenev}},\ and\ \bibinfo
  {author} {\bibfnamefont {L.~I.}\ \bibnamefont {Glazman}},\ }\bibfield
  {title} {\bibinfo {title} {{Dynamic Response of One-Dimensional Interacting
  Fermions}},\ }\href {https://link.aps.org/doi/10.1103/PhysRevLett.96.196405}
  {\bibfield  {journal} {\bibinfo  {journal} {Phys. Rev. Lett.}\ }\textbf
  {\bibinfo {volume} {96}},\ \bibinfo {pages} {196405} (\bibinfo {year}
  {2006})}\BibitemShut {NoStop}%
\bibitem [{\citenamefont {Pereira}\ \emph {et~al.}(2008)\citenamefont
  {Pereira}, \citenamefont {White},\ and\ \citenamefont
  {Affleck}}]{Pereira:2008bw}%
  \BibitemOpen
  \bibfield  {author} {\bibinfo {author} {\bibfnamefont {R.~G.}\ \bibnamefont
  {Pereira}}, \bibinfo {author} {\bibfnamefont {S.~R.}\ \bibnamefont {White}},\
  and\ \bibinfo {author} {\bibfnamefont {I.}~\bibnamefont {Affleck}},\
  }\bibfield  {title} {\bibinfo {title} {{Exact Edge Singularities and
  Dynamical Correlations in Spin-1/2 Chains}},\ }\href
  {https://link.aps.org/doi/10.1103/PhysRevLett.100.027206} {\bibfield
  {journal} {\bibinfo  {journal} {Phys. Rev. Lett.}\ }\textbf {\bibinfo
  {volume} {100}},\ \bibinfo {pages} {027206} (\bibinfo {year}
  {2008})}\BibitemShut {NoStop}%
\bibitem [{\citenamefont {Imambekov}\ and\ \citenamefont
  {Glazman}(2009)}]{Imambekov:2009ly}%
  \BibitemOpen
  \bibfield  {author} {\bibinfo {author} {\bibfnamefont {A.}~\bibnamefont
  {Imambekov}}\ and\ \bibinfo {author} {\bibfnamefont {L.~I.}\ \bibnamefont
  {Glazman}},\ }\bibfield  {title} {\bibinfo {title} {{P}henomenology of
  {O}ne-{D}imensional {Q}uantum {L}iquids {B}eyond the {L}ow-{E}nergy
  {L}imit},\ }\href {https://doi.org/10.1103/physrevlett.102.126405} {\bibfield
   {journal} {\bibinfo  {journal} {Phys. Rev. Lett.}\ }\textbf {\bibinfo
  {volume} {102}},\ \bibinfo {pages} {126405} (\bibinfo {year}
  {2009})}\BibitemShut {NoStop}%
\bibitem [{\citenamefont {Olshanii}(1998)}]{Olshanii:1998cf}%
  \BibitemOpen
  \bibfield  {author} {\bibinfo {author} {\bibfnamefont {M.}~\bibnamefont
  {Olshanii}},\ }\bibfield  {title} {\bibinfo {title} {{A}tomic {S}cattering in
  the {P}resence of an {E}xternal {C}onfinement and a {G}as of {I}mpenetrable
  {B}osons},\ }\href {https://doi.org/10.1103/physrevlett.81.938} {\bibfield
  {journal} {\bibinfo  {journal} {Phys. Rev. Lett.}\ }\textbf {\bibinfo
  {volume} {81}},\ \bibinfo {pages} {938} (\bibinfo {year} {1998})}\BibitemShut
  {NoStop}%
\bibitem [{\citenamefont {Sigma-Aldrich}(2008)}]{mcm41SA:2008}%
  \BibitemOpen
  \bibfield  {author} {\bibinfo {author} {\bibnamefont {Sigma-Aldrich}},\
  }\bibfield  {title} {\bibinfo {title} {{Synthesis of Mesoporous Materials}},\
  }\href
  {https://www.sigmaaldrich.com/technical-documents/articles/material-matters/mesoporous-materials.html}
  {\bibfield  {journal} {\bibinfo  {journal} {Mater. Matters}\ }\textbf
  {\bibinfo {volume} {3.1}},\ \bibinfo {pages} {17} (\bibinfo {year}
  {2008})}\BibitemShut {NoStop}%
\bibitem [{\citenamefont {Brunauer}\ \emph {et~al.}(1938)\citenamefont
  {Brunauer}, \citenamefont {Emmett},\ and\ \citenamefont
  {Teller}}]{Brunauer:1938pz}%
  \BibitemOpen
  \bibfield  {author} {\bibinfo {author} {\bibfnamefont {S.}~\bibnamefont
  {Brunauer}}, \bibinfo {author} {\bibfnamefont {P.~H.}\ \bibnamefont
  {Emmett}},\ and\ \bibinfo {author} {\bibfnamefont {E.}~\bibnamefont
  {Teller}},\ }\bibfield  {title} {\bibinfo {title} {{A}dsorption of {G}ases in
  {M}ultimolecular {L}ayers},\ }\href
  {https://pubs.acs.org/doi/abs/10.1021/ja01269a023} {\bibfield  {journal}
  {\bibinfo  {journal} {J. Am. Chem. Soc.}\ }\textbf {\bibinfo {volume} {60}},\
  \bibinfo {pages} {309} (\bibinfo {year} {1938})}\BibitemShut {NoStop}%
\bibitem [{\citenamefont {Jaroniec}\ \emph {et~al.}(1999)\citenamefont
  {Jaroniec}, \citenamefont {Kruk},\ and\ \citenamefont
  {Olivier}}]{Jaroniec:1999mi}%
  \BibitemOpen
  \bibfield  {author} {\bibinfo {author} {\bibfnamefont {M.}~\bibnamefont
  {Jaroniec}}, \bibinfo {author} {\bibfnamefont {M.}~\bibnamefont {Kruk}},\
  and\ \bibinfo {author} {\bibfnamefont {J.~P.}\ \bibnamefont {Olivier}},\
  }\bibfield  {title} {\bibinfo {title} {{S}tandard {N}itrogen {A}dsorption
  {D}ata for {C}haracterization of {N}anoporous {S}ilicas},\ }\href
  {https://pubs.acs.org/doi/10.1021/la990136e} {\bibfield  {journal} {\bibinfo
  {journal} {Langmuir}\ }\textbf {\bibinfo {volume} {15}},\ \bibinfo {pages}
  {5410} (\bibinfo {year} {1999})}\BibitemShut {NoStop}%
\bibitem [{\citenamefont {Copley}\ and\ \citenamefont
  {Cook}(2003)}]{Copley:2003dc}%
  \BibitemOpen
  \bibfield  {author} {\bibinfo {author} {\bibfnamefont {J.}~\bibnamefont
  {Copley}}\ and\ \bibinfo {author} {\bibfnamefont {J.}~\bibnamefont {Cook}},\
  }\bibfield  {title} {\bibinfo {title} {{The Disk Chopper Spectrometer at
  {NIST}: a new instrument for quasielastic neutron scattering studies}},\
  }\href {https://doi.org/10.1016/s0301-0104(03)00124-1} {\bibfield  {journal}
  {\bibinfo  {journal} {Chem. Phys.}\ }\textbf {\bibinfo {volume} {292}},\
  \bibinfo {pages} {477} (\bibinfo {year} {2003})}\BibitemShut {NoStop}%
\bibitem [{\citenamefont {Azuah}\ \emph {et~al.}(2009)\citenamefont {Azuah},
  \citenamefont {Kneller}, \citenamefont {Qiu}, \citenamefont
  {Tregenna-Piggott}, \citenamefont {Brown}, \citenamefont {Copley},\ and\
  \citenamefont {Dimeo}}]{Azuah:2009cs}%
  \BibitemOpen
  \bibfield  {author} {\bibinfo {author} {\bibfnamefont {R.~T.}\ \bibnamefont
  {Azuah}}, \bibinfo {author} {\bibfnamefont {L.~R.}\ \bibnamefont {Kneller}},
  \bibinfo {author} {\bibfnamefont {Y.}~\bibnamefont {Qiu}}, \bibinfo {author}
  {\bibfnamefont {P.~L.~W.}\ \bibnamefont {Tregenna-Piggott}}, \bibinfo
  {author} {\bibfnamefont {C.~M.}\ \bibnamefont {Brown}}, \bibinfo {author}
  {\bibfnamefont {J.~R.~D.}\ \bibnamefont {Copley}},\ and\ \bibinfo {author}
  {\bibfnamefont {R.~M.}\ \bibnamefont {Dimeo}},\ }\bibfield  {title} {\bibinfo
  {title} {{D{A}VE}: {A} comprehensive {S}oftware {S}uite for the {R}eduction,
  {V}isualization, and {A}nalysis of {L}ow {E}nergy {N}eutron {S}pectroscopic
  {D}ata},\ }\href {https://dx.doi.org/10.6028/jres.114.025} {\bibfield
  {journal} {\bibinfo  {journal} {J. Res. Natl. Inst. Stand. Technol.}\
  }\textbf {\bibinfo {volume} {114}},\ \bibinfo {pages} {341} (\bibinfo {year}
  {2009})}\BibitemShut {NoStop}%
\bibitem [{\citenamefont {Aziz}\ \emph {et~al.}(1979)\citenamefont {Aziz},
  \citenamefont {Nain}, \citenamefont {Carley}, \citenamefont {Taylor},\ and\
  \citenamefont {McConville}}]{Aziz:1979hs}%
  \BibitemOpen
  \bibfield  {author} {\bibinfo {author} {\bibfnamefont {R.~A.}\ \bibnamefont
  {Aziz}}, \bibinfo {author} {\bibfnamefont {V.~P.~S.}\ \bibnamefont {Nain}},
  \bibinfo {author} {\bibfnamefont {J.~S.}\ \bibnamefont {Carley}}, \bibinfo
  {author} {\bibfnamefont {W.~L.}\ \bibnamefont {Taylor}},\ and\ \bibinfo
  {author} {\bibfnamefont {G.~T.}\ \bibnamefont {McConville}},\ }\bibfield
  {title} {\bibinfo {title} {{An accurate intermolecular potential for
  helium}},\ }\href {http://dx.doi.org/10.1063/1.438007} {\bibfield  {journal}
  {\bibinfo  {journal} {J. Chem. Phys.}\ }\textbf {\bibinfo {volume} {70}},\
  \bibinfo {pages} {4330} (\bibinfo {year} {1979})}\BibitemShut {NoStop}%
\bibitem [{\citenamefont {Przybytek}\ \emph {et~al.}(2010)\citenamefont
  {Przybytek}, \citenamefont {Cencek}, \citenamefont {Komasa}, \citenamefont
  {{\L}ach}, \citenamefont {Jeziorski},\ and\ \citenamefont
  {Szalewicz}}]{Przybytek:2010js}%
  \BibitemOpen
  \bibfield  {author} {\bibinfo {author} {\bibfnamefont {M.}~\bibnamefont
  {Przybytek}}, \bibinfo {author} {\bibfnamefont {W.}~\bibnamefont {Cencek}},
  \bibinfo {author} {\bibfnamefont {J.}~\bibnamefont {Komasa}}, \bibinfo
  {author} {\bibfnamefont {G.}~\bibnamefont {{\L}ach}}, \bibinfo {author}
  {\bibfnamefont {B.}~\bibnamefont {Jeziorski}},\ and\ \bibinfo {author}
  {\bibfnamefont {K.}~\bibnamefont {Szalewicz}},\ }\bibfield  {title} {\bibinfo
  {title} {{Relativistic and Quantum Electrodynamics Effects in the Helium Pair
  Potential}},\ }\href {http://link.aps.org/doi/10.1103/PhysRevLett.104.183003}
  {\bibfield  {journal} {\bibinfo  {journal} {Phys. Rev. Lett.}\ }\textbf
  {\bibinfo {volume} {104}},\ \bibinfo {pages} {183003} (\bibinfo {year}
  {2010})}\BibitemShut {NoStop}%
\bibitem [{\citenamefont {Cencek}\ \emph {et~al.}(2012)\citenamefont {Cencek},
  \citenamefont {Przybytek}, \citenamefont {Komasa}, \citenamefont {Mehl},
  \citenamefont {Jeziorski},\ and\ \citenamefont {Szalewicz}}]{Cencek:2012iz}%
  \BibitemOpen
  \bibfield  {author} {\bibinfo {author} {\bibfnamefont {W.}~\bibnamefont
  {Cencek}}, \bibinfo {author} {\bibfnamefont {M.}~\bibnamefont {Przybytek}},
  \bibinfo {author} {\bibfnamefont {J.}~\bibnamefont {Komasa}}, \bibinfo
  {author} {\bibfnamefont {J.~B.}\ \bibnamefont {Mehl}}, \bibinfo {author}
  {\bibfnamefont {B.}~\bibnamefont {Jeziorski}},\ and\ \bibinfo {author}
  {\bibfnamefont {K.}~\bibnamefont {Szalewicz}},\ }\bibfield  {title} {\bibinfo
  {title} {{Effects of adiabatic, relativistic, and quantum electrodynamics
  interactions on the pair potential and thermophysical properties of
  helium}},\ }\href
  {http://scitation.aip.org/content/aip/journal/jcp/136/22/10.1063/1.4712218}
  {\bibfield  {journal} {\bibinfo  {journal} {J. Chem. Phys.}\ }\textbf
  {\bibinfo {volume} {136}},\ \bibinfo {pages} {224303} (\bibinfo {year}
  {2012})}\BibitemShut {NoStop}%
\bibitem [{\citenamefont {Ceperley}(1995)}]{Ceperley:1995gr}%
  \BibitemOpen
  \bibfield  {author} {\bibinfo {author} {\bibfnamefont {D.~M.}\ \bibnamefont
  {Ceperley}},\ }\bibfield  {title} {\bibinfo {title} {{Path integrals in the
  theory of condensed helium}},\ }\href
  {http://link.aps.org/doi/10.1103/RevModPhys.67.279} {\bibfield  {journal}
  {\bibinfo  {journal} {Rev. Mod. Phys.}\ }\textbf {\bibinfo {volume} {67}},\
  \bibinfo {pages} {279} (\bibinfo {year} {1995})}\BibitemShut {NoStop}%
\bibitem [{\citenamefont {Boninsegni}\ \emph {et~al.}(2006)\citenamefont
  {Boninsegni}, \citenamefont {{Prokof'ev}},\ and\ \citenamefont
  {Svistunov}}]{Boninsegni:2006ed}%
  \BibitemOpen
  \bibfield  {author} {\bibinfo {author} {\bibfnamefont {M.}~\bibnamefont
  {Boninsegni}}, \bibinfo {author} {\bibfnamefont {N.}~\bibnamefont
  {{Prokof'ev}}},\ and\ \bibinfo {author} {\bibfnamefont {B.}~\bibnamefont
  {Svistunov}},\ }\bibfield  {title} {\bibinfo {title} {{Worm Algorithm for
  Continuous-Space Path Integral Monte Carlo Simulations}},\ }\href
  {https://link.aps.org/doi/10.1103/PhysRevLett.96.070601} {\bibfield
  {journal} {\bibinfo  {journal} {Phys. Rev. Lett.}\ }\textbf {\bibinfo
  {volume} {96}},\ \bibinfo {pages} {070601} (\bibinfo {year}
  {2006})}\BibitemShut {NoStop}%
\bibitem [{\citenamefont {Suzuki}(1990)}]{Suzuki1990}%
  \BibitemOpen
  \bibfield  {author} {\bibinfo {author} {\bibfnamefont {M.}~\bibnamefont
  {Suzuki}},\ }\bibfield  {title} {\bibinfo {title} {Fractal decomposition of
  exponential operators with applications to many-body theories and monte carlo
  simulations},\ }\href
  {https://doi.org/https://doi.org/10.1016/0375-9601(90)90962-N} {\bibfield
  {journal} {\bibinfo  {journal} {Phys. Lett. A}\ }\textbf {\bibinfo {volume}
  {146}},\ \bibinfo {pages} {319 } (\bibinfo {year} {1990})}\BibitemShut
  {NoStop}%
\bibitem [{\citenamefont {{Del Maestro}}(2022)}]{delmaestro:code}%
  \BibitemOpen
  \bibfield  {author} {\bibinfo {author} {\bibfnamefont {A.}~\bibnamefont {{Del
  Maestro}}},\ }\bibfield  {title} {\bibinfo {title} {Path integral quantum
  monte carlo},\ }\href {https://code.delmaestro.org} {\bibfield  {journal}
  {\bibinfo  {journal} {{Online}}\ } (\bibinfo {year} {2022})},\ \bibinfo
  {note} {\url{https://code.delmaestro.org}}\BibitemShut {NoStop}%
\bibitem [{\citenamefont {Del~Maestro}(2012)}]{DelMaestro:2012ba}%
  \BibitemOpen
  \bibfield  {author} {\bibinfo {author} {\bibfnamefont {A.}~\bibnamefont
  {Del~Maestro}},\ }\bibfield  {title} {\bibinfo {title} {{A Luttinger Liquid
  Core Inside Helium-4 Filled Nanopores}},\ }\href
  {http://www.worldscientific.com/doi/abs/10.1142/S021797921244002X} {\bibfield
   {journal} {\bibinfo  {journal} {Int. J. Mod. Phys. B}\ }\textbf {\bibinfo
  {volume} {26}},\ \bibinfo {pages} {1244002} (\bibinfo {year}
  {2012})}\BibitemShut {NoStop}%
\bibitem [{rep(2022)}]{repo}%
  \BibitemOpen
  \href {https://doi.org/10.5281/zenodo.6112399} {} (\bibinfo {year} {2022}),\
  \bibinfo {note} {{All code, scripts and data used in this work are included
  in a GitHub repository:
  \url{https://github.com/DelMaestroGroup/papers-code-preplated-nanopores-scattering},
  Permanent link: \url{https://doi.org/10.5281/zenodo.6112399}}}\BibitemShut
  {NoStop}%
\end{thebibliography}%
\FloatBarrier

\acknowledgments

We acknowledge the support of the National Institute of Standards and Technology, U.S. Department of Commerce, in providing the neutron research facilities used in this work.  This research was supported in part by the National Science Foundation (NSF) under award Nos.~DMR-1809027 and DMR-1808440.  This work used the Extreme Science and Engineering Discovery Environment (XSEDE), which is supported by NSF grant number ACI-1548562. XSEDE resources used include Bridges at Pittsburgh Supercomputing, Comet at San Diego Supercomputer Center, and Open Science Grid (OSG) through allocations TG-DMR190045 and TG-DMR190101.  OSG is supported by the NSF under award No.~1148698, and the U.S. Department of Energy's Office of Science.  Certain commercial equipment, instruments, or materials (or suppliers, or software, $\dots$) are identified in this paper to foster understanding. Such identification does not imply recommendation or endorsement by the National Institute of Standards and Technology, nor does it imply that the materials or equipment identified are necessarily the best available for the purpose.




\section*{Methods}

\section{Sample Characterization}
MCM-41 is a mesoporous material with a hierarchical structure consisting of cylindrical pores arranged in a hexagonal lattice. It is produced using a surfactant templating technique that produces pores that are monodisperse, unidirectional, and have a regular 2D hexagonal structure.  The typical aspect ratio of the pores is $\sim 1000:1$ making them an attractive medium for studies of 1D behavior. The sample and characterization techniques have been reported previously and we will briefly review them here.    

Our sample was obtained from Sigma-Aldrich \cite{mcm41SA:2008} and was characterized using X-ray powder diffraction and N$_2$ gas adsorption isotherm measurements \cite{Prisk:2013cu}. The X-ray diffraction data indicated that the sample consisted of a single phase with pores arranged on a hexagonal lattice with a lattice constant of \SI{4.7}{\nano\meter}.  A Brunauer-Emmett-Teller (BET) analysis \cite{Brunauer:1938pz} of the N$_2$ isotherm gave a surface area of \SI{915}{\meter^2\per\gram}.  The pore diameter size distribution was calculated using the Kruk-Jaroniec-Sayari method \cite{Jaroniec:1999mi} and was found to be Gaussian with a mean value of \SI{3.0}{\nano\meter} and a full-width at half-maximum of \SI{0.3}{\nano\meter}. 

Adsorption isotherms were also carried out with research grade Ar gas  at \SI{90}{\kelvin} to determine the monolayer coverage \cite{Nichols:2020of}. A BET analysis of the isotherm yielded a monolayer coverage of \SI{8.994}{\milli\mol\per\gram}. This monolayer coverage, when combined with the measured surface area, yields an aerial coverage of $\SI{0.59}{\angstrom^{-2}}$ and, using the van der Waals radius for Ar, a monolayer density of $n_{\rm Ar} = \SI{0.017}{\angstrom^{-3}}$.

We also carried out $^4$He isotherms on MCM-41 preplated with a single monolayer of Ar.  The Ar pre-plating was carried out at \SI{90}{\kelvin} and then the sample was slowly cooled to \SI{4.2}{\kelvin} over the course of several hours.  $^4$He isotherms were then carried out at \SI{4.2}{\kelvin} using standard volumetric techniques.  The results are shown in Figure 1 of the main text.  The initially adsorbed $^4$He is strongly bound to the surface of the MCM-41 resulting in zero pressure rise until $\sim \SI{7.5}{\milli\mol\per\gram}$ has been adsorbed.  There is a small region between $\sim\SI{7.5}{\milli\mol\per\gram}$ and \SI{13}{\milli\mol\per\gram}  where the pressure increases.  Once a  filling of \SI{13}{\milli\mol\per\gram} has been reached no additional helium is adsorbed into the pores until the pressure is close to the bulk vapor pressure $P_0$.  Once $P/P_0$ is greater than $\sim 0.9$, $^4$He capillary condenses between the MCM-41 grains.

\section{Neutron Scattering}
Neutron scattering studies of $^4$He in Ar preplated MCM-41 were carried out using the Disc Chopper Spectrometer (DCS) at the NIST Center for Neutron Research \cite{Copley:2003dc}. This instrument is a direct geometry time-of-flight chopper spectrometer which views a cold moderator. High speed choppers are used to create a pulsed neutron beam with a well defined incident wavelength. Neutrons scattered by the sample are detected by a secondary spectrometer consisting of 913 $^3$He detectors \SI{4.01}{\meter} from the sample at scattering angles from \SI{5}{\degree} to \SI{140}{\degree}. Standard data reduction routines \cite{Azuah:2009cs} are used to convert the observed scattering to the dynamic structure factor $S(Q,E)$.

The sample for these studies consisted of \SI{6.13}{\gram} of MCM-41 inside a cylindrical aluminum can.  The MCM-41 was in the form of cylindrical pellets \SI{1.25}{\centi\meter} in diameter and \SI{1}{\centi\meter} high with a mass of \SI{0.875}{\gram}.  The pellets were baked in vacuum at \SI{120}{\celsius} to remove adsorbed water vapor.  The sample was then transferred to an aluminum sample cell in a nitrogen glove box.  The cell was a cylindrical aluminum can of outer diameter \SI{1.5}{\centi\meter}, a height of \SI{6}{\centi\meter}, and a wall thickness of \SI{1}{\milli\meter}. The pellets were separated by cadmium spacers to reduce multiple scattering. A top-loading liquid helium cryostat with aluminum tails, commonly referred to as an ``orange" cryostat, was used to obtain the low temperatures examined in this study. A silicon diode was used to monitor the sample temperature.

Incident wavevectors of \SI{4.0}, \SI{2.5} and \SI{1.71}{\angstrom^{-1}} were used (see Figure 4 in the main text for incident wavevector dependence).  Longer incident wavelengths offer the advantage of better energy resolution but at significantly decreased intensity and range of momentum transfers, $Q$, accessible.  The studies at \SI{4.0}{\angstrom^{-1}} had an energy resolution of \SI{93.6}{\micro\eV}.  However, the flux was limited to $1.05\times 10^5$~neutrons/cm${^2}$/s and the maximum $Q$ was \SI{2.9}{\angstrom^{-1}}. Due to this lower flux and limited Q range we limited these studies to two fillings of monolayer and slightly overfilled pores.  More extensive studies at a variety of fillings were carried out at \SI{2.5}{\angstrom^{-1}}.  The energy resolution was significantly larger (\SI{772}{\micro\eV}) but with much larger incident flux ($8.75\times10^5$ neutrons/cm$^2$/s) and Q range (\SI{4.6}{\angstrom^{-1}}). A limited number of measurements were also carried out at \SI{1.71}{\angstrom^{-1}}. These measurements had significantly worse energy resolution (\SI{2370}{\micro\eV}) but a much expanded Q range (\SI{6.8}{\angstrom^{-1}}).  However, since DCS is located on a cold moderator the flux is significantly decreased ($1.89\times 10^5$ neutrons/cm$^2$/s) at these short wavelengths.

\section{Theoretical Modeling}

An \emph{ab initio} model of superfluid ${^4}$He confined inside ordered nanoporous MCM-41 pre-pated with argon gas can be constructed from the superposition of single pores, each described by a $N$-body Hamiltonian:
\begin{equation}
    H = -\frac{\hbar^2}{2m}\sum_{i=1}^N \nabla_i^2 + \sum_{i=1}^N U_{\rm pore}(\vec{r}_i) + \frac{1}{2}\sum_{i,j} V_{\rm He}(\vec{r_i}-\vec{r_j})
\label{eq:Ham}
\end{equation}
where $m$ is the mass of a single ${^4}$He atom located at position $\vec{r}_i = (x_i,y_i,z_i)$ confined inside a nanopore by $U_{\rm pore}$ and interacting with other He atoms through $V_{\rm He}$.  Both potential energy terms arise from induced dipole-dipole interactions.  $U_{\rm pore}$ 
was recently determined for the specific system under consideration here \cite{Nichols:2020of} while $V_{\rm He}$ is known to high precision \cite{Aziz:1979hs, Przybytek:2010js, Cencek:2012iz}.

\subsection{Quantum Monte Carlo Method}
\label{subsec:QMC}

A system of confined helium described by Eq.~(\ref{eq:Ham}) was simulated using a quantum Monte Carlo algorithm exploiting path integrals \cite{Ceperley:1995gr,Boninsegni:2006ed,Nichols:2020of}.   $T>0$ expectation values of observables $\mathcal{O}$ were sampled via 
\begin{equation}
    \expval{{\mathcal{O}}}  = \frac{1}{\mathcal{Z}} \Tr{\mathcal{O}\ \mathrm{e}^{-\beta {{H}}}}
\label{eq:thermal_expectation_value}
\end{equation}
where $\beta = 1/k_{\rm {B}} T$ is the inverse temperature and the partition function $\mathcal{Z} = \Tr\mathrm{e}^{-\beta {{H}}}$ can be written as a sum of discrete imaginary time paths (worldlines) over the set of all permutations $\mathcal{P}$ of the first quantized labels of the $N$ indistinguishable $^4$He atoms: 
\begin{equation}
    \mathcal{Z} \simeq
            \frac{1}{\qty(4\Lambda \tau)^{3NM/2}}\frac{1}{N!}\sum_{\mathcal{P}}  \qty[\prod_{\alpha=0}^{M-1} \int \mathcal{D} R_{\alpha}] \mathrm{e}^{-\sum_{\alpha=0}^{M-1}\mathcal{S}_{\alpha}}\, .
\label{eq:Z}
\end{equation}
Here $\tau = \beta/M$ is the imaginary time step where  $M \in \mathbb{Z} \gg 1$ and $R_\alpha \equiv \boldsymbol{r}_{\alpha,1},\boldsymbol{r}_{\alpha,2},\dots,\boldsymbol{r}_{\alpha,N}$ are the spatial positions of the particles at imaginary time slice $\alpha$. Bosonic symmetry restricts $\mathcal{P}R_{M} = R_{0}$ and we have employed the short-hand notation $\int \mathcal{D} R_\alpha \equiv \prod_{i=1}^{N} \int \dd{\vb*{r}_{\alpha,i}}$ and $\Lambda = {\hbar^2}/({2m})$.
Discrete imaginary time-step errors are suppressed to $\mathrm{O}(\tau^4)$ \cite{Suzuki1990} through an effective imaginary time action
\begin{equation}\begin{aligned}
    \mathcal{S}_\alpha =  \sum_{i=1}^N \frac{\norm{\vb{r}_{\alpha,i}-\vb{r}_{\alpha+1,i}}^2}{4\Lambda \tau}
+ \tau\qty[1+\tfrac{1}{3}(-1)^\alpha]\mathcal{V}(R_\alpha) \\ 
+ \;  \tau^3\qty[1-(-1)^\alpha] \frac{\Lambda}{9}\sum_{i=1}^{N}\norm{\nabla_i \mathcal{V}(R_\alpha)}^2 
\label{eq:Salpha}
\end{aligned}\end{equation}
where 
\begin{equation}
    \mathcal{V}(R_\alpha) \equiv \sum_{i=1}^N {U}_{\rm pore}(\vb*{r}_{\alpha,i}) + \sum_{i<j}{V}_{\rm He}(\vb*{r}_{\alpha,i}-\vb*{r}_{\alpha,j})\ . 
\label{eq:potentialdef}
\end{equation}

Simulations were performed at $T = \SI{1.6}{\kelvin}$ for four chemical potentials: $\mu / k_{\rm B} = $ \SIlist{-47;-27;-19;-7}{\kelvin} that were identified as representative of the nanopores at different stages of filling-- from a single adsorbed layer at $\mu / k_{\rm B} = \SI{-47}{\kelvin}$ to a fully filled pore at $\mu / k_{\rm B} = \SI{-7}{\kelvin}$ (see Fig.~1 in the main text). The pores had outer radius $R = \SI{15.51}{\angstrom}$ and length $L = \SI{50}{\angstrom}$ and all simulations were performed in the grand canonical ensemble yielding $N \approx 600$ helium atoms for $\mu / k_{\rm B} = \SI{-7}{\kelvin}$.  Trotter errors were deemed to be smaller than statistical uncertainties for $\tau \cdot k_{\rm B} = \SI{1/250}{\kelvin^{-1}}$, which was used for all simulations.  Further details (including the effects of the finite pore length) are reported in Ref.~\cite{Nichols:2020of}. The quantum Monte Carlo software used to produce all results is available online \cite{delmaestro:code}.

\subsection{Observables}
Using the quantum Monte Carlo method described above we measure the radial density (Figure 1): 
\begin{equation}
    \rho_{\rm rad}(r) = \left \langle \sum_{i=1}^{N} \delta\qty(\sqrt{x_i^2 + y_i^2} -r) \right \rangle \, ,
\label{eq:radial_density}
\end{equation}
the density-density correlation function (Figure 2 \textbf{b},\textbf{d}):
\begin{equation}
    g_2(r) = \expval{\rho(r)\rho(0)} = \expval{\frac{V}{N^2} \sum_{i\ne j} \delta\qty(r - \abs{\vb*{r}_i - \vb*{r}_j})}\, ,
\label{eq:g2}
\end{equation}
and the structure factor (elastic scattering, Figure 2 \textbf{c},\textbf{e}):
\begin{equation}
    S(\vb*{Q}) = \expval{\frac{1}{N} \rho(\vb*{Q}) \rho(-\vb*{Q})}\, .
    \label{eq:SQ}
\end{equation}
Here $V$ is the accessible volume of the pore and $\rho(\vb*{Q}) = \sum_{i=1}^{N} \exp(-i\vb*{Q}\cdot \vb*{r}_i)$. The \emph{core} versions of these estimators in Figure 2\textbf{d} -- \textbf{e} include only those $^4$He atoms with $\sqrt{x_i^2 + y_i^2} < \SI{1.72}{\angstrom}$ identified as the location of the first minimum of $\rho_{\rm rad} (r)$ shown in Figure 1\textbf{b} for the full pore with $\mu / k_{\rm B} = \SI{-7}{\kelvin}$.

For the inelastic scattering $S(Q,E)$ in Figure 3\textbf{b}, simulations were performed on a purely one-dimensional system of $^4$He with $L=\SI{200}{\angstrom}$ in the canonical ensemble at fixed $\rho_{1D} = \SI{0.14}{\angstrom^{-1}}$ and $T = \SI{1.6}{\kelvin}$.  The intermediate scattering function:
\begin{equation}
    F(\vb*{Q},\alpha\tau) = \frac{1}{N}\expval{ \sum_{j,k} e^{-i\vb*{Q}\cdot\vb*{r}_{\alpha,j}} e^{i\vb*{Q}\cdot\vb*{r}_{0,k}}}
\end{equation}
is related to the dynamic structure factor $S(\vb*{Q},E)$ via:
\begin{equation}
    F(\vb*{Q},\tau) = \frac{1}{\hbar}\int_0^\infty S(\vb*{Q},E) \qty[e^{-\tau E} + e^{-(\beta - \tau)E}] \dd{E} 
\label{eq:intermediate_scattering_function}
\end{equation}
which can be inverted using a recently introduced parameter-free differential evolution algorithm for imaginary time correlation functions \cite{Nichols:2022d}. 
\vspace{0.5em}

\subsection{Tomonaga-Luttinger Liquid Predictions}

The one-dimensional density-density correlation function can be computed within the effective quantum hydrodynamic theory to be \cite{DelMaestro:2011dh,DelMaestro:2012ba}:
\begin{multline}
\label{eq:rhoz1D}
\langle\rho(z) \rho(0)\rangle=  \rho_{1D}^{2}+\frac{K}{2 \pi^{2}} \frac{d^{2}}{d z^{2}} \ln \theta_{1}\left[\frac{\pi z}{L}, e^{-\pi \hbar v / L k_{\rm B} T}\right]  \\
\quad +\mathcal{A} \cos \left(2 \pi \rho_{1D} z\right)\left\{\frac{2 \eta\left(\frac{i \hbar v}{L k_{\rm B} T}\right) e^{-\pi \hbar v / 6 L k_{\rm B} T}}{\theta_{1}\left(\frac{\pi z}{L}, e^{-\pi \hbar v / L k_{\rm B}T}\right)}\right\}^{2 K}
\end{multline}
where $\eta(\cdot)$ is the Dedekind eta function and $\theta_1(z,q)$ is the Jacobi theta function of the first kind. This expression depends on 3 parameters: $K,v$ and $\mathcal{A}$ with $\rho_{1D}$ being determined from simulations as $\rho_{1D}=N/L$. In practice it is beneficial to first fit the envelope of decay from the asymptotic form:
\begin{equation}
\langle\rho(z) \rho(0)\rangle 
\underset{T\to 0}{\overset{L\to\infty}{\approx}}
\rho_{0}^{2}-\frac{K}{2 \pi^{2} z^{2}}+\frac{\mathcal{A}}{z^{2K }} \cos \left(2 \pi \rho_{0} z\right)
\end{equation}
to obtain reasonable initial values for the parameters before performing a full non-linear fit to Eq.~\eqref{eq:rhoz1D}. For the core $^4$He atoms with $\rho_{1D} = 0.293(2)$ shown in Figure 2\textbf{d} we find $K = 0.15(4)$, $\hbar v / k_{B} =8(3)~\si{\angstrom \kelvin}$ and $\mathcal{A} = 0.036(2)$. The different value of $K$ extracted in this simulation compared to that found by analyzing the experimentally determined $S(Q,E)$ is due to this pore being closer to the fully filled regime with $\rho_{1D} \approx 0.3$ as opposed to $\rho_{1D} \approx 0.25$ found in the experiment.

\section{Data Availability}
All experimental and quantum Monte Carlo data, code, and scripts used to generate all figures in this paper are available online \cite{repo}.

\ifarXiv
    

\section*{Supplementary material}

# Supplementary Information for "Experimental Realization of One Dimensional Helium"

Adrian Del Maestro, Nathan S. Nichols, Timothy R. Prisk, Garfield Warren, Paul E. Sokol

## ELASTIC SCATTERING

Experimental adsorption isotherms were used to connect the experimental results to quantum Monte Carlo calculations of the layer density in the pores [S1]. As noted in the main text, pore completion occurs at 13 mmol/g and this value was used to normalize to the results of Nichols *et al.* [S1] to provide filling ranges for each of the layers. These computed values are provided in Table S1.

| Layer | Filling (mmol/g) | Fraction of atoms | Spacing (Å) |
| --- | --- | --- | --- |
| Layer 1 | 0.00 - 6.53 | 0.50 | 3.16 |
| Layer 2 | 6.53 - 10.73 | 0.32 | 3.24 |
| Layer 3 | 10.73 - 12.71 | 0.15 | 3.53 |
| Pore Center | 12.73 - 13.01 | 0.02 | 3.9 |

TABLE S1. **Layer Filling Properties.** The range of fillings (in mmol/g), fraction of atoms, and inter-layer spacing for each of the layers. The filling scale is obtained by scaling the full pore filling, as determined from adsorption isotherms, to quantum Monte Carlo calculations.

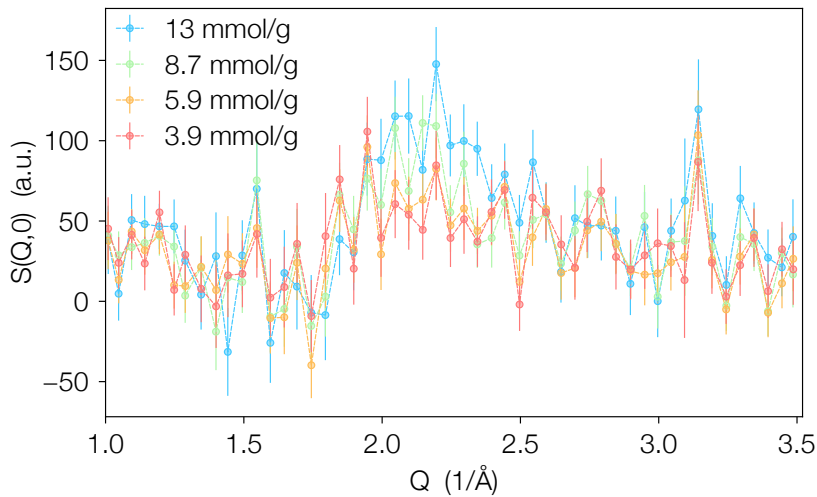

FIG. S1. **Elastic scattering from confined helium.** Experimental measurements of $S(Q, 0)$ at fillings of 3.89 (red), 5.91 (orange), 8.68 (green) and 13.0 (blue ) mmol/g corresponding roughly to those filling fractions where layer completion may occur. Measurements were carried out with an incident wavelength of $2.5\,\text{Å}^{-1}$ and the scattering from the Ar plated MCM-41 has been subtracted. Note the agreement with Figure 2**c** in the main text computed from quantum Monte Carlo simulations.

As can be seen in Table S1, half of the atoms reside in the first layer and are strongly bound to the Ar/MCM-41 surface. Layers 2 and 3, which are less strongly bound but still solid like, account for an additional 47% of the material in the pores. Finally, the central core liquid, since it is only 1 atom in lateral extent, accounts for only 2-3% of the helium in the pores.

The elastic scattering with helium fillings of 3.89, 5.91, 8.68, and 13.0 mmol/g is shown in Figure S1. Figure 2**a** in the main text is computed from performing subtractions within this data set and we note the strong similarity with the simulation results presented in Figure 2**c** of the main text. The two lower fillings correspond to a nearly completed monolayer and nearly completed second layer. The scattering is strong at small $Q$ then rapidly decreases in intensity with increasing $Q$ due to grain size of the MCM-41. There is also a relatively broad peak at $\sim 2.1\,\text{Å}^{-1}$

and a relatively constant scattering at higher $Q$. We attribute this scattering to the solid layers that are forming near the pore wall at larger filling. The peak corresponds to an interatomic spacing of $\sim 3$ Å consistent with the values in Table S1. We also note that the increase in intensity of the 8.68 mmol/g results with respect to those for 5.91 mmol/g is consistent with the expectation that the monolayer contains 50% of the atoms in the full pore while the second layer only contributes and additional 30%.

We note that the peak at $1.6\,\text{Å}^{-1}$ is present but does not stand out clearly in these plots. One must take into consideration that the core liquid represents only 2% of the total scattering while layers 1-3 represent 98% of the intensity. Also, a remnant of the peak is still visible at 8.68 mmol/g. We believe this is due to the fact that this sequence of measurements was carried out upon desorption of $^4$He from the pores and that some hysteresis occurred leaving a small amount of core liquid still present.

[S1] N. S. Nichols, T. R. Prisk, G. Warren, P. Sokol, and A. Del Maestro, Dimensional reduction of helium-4 inside argon-plated MCM-41 nanopores, Phys. Rev. B **102**, 144505 (2020).


\fi

\end{document}